\documentclass[pdflatex,sn-mathphys-num]{sn-jnl}

\usepackage{graphicx}%
\usepackage{multirow}%
\usepackage{amsmath,amssymb,amsfonts}%
\usepackage{amsthm}%
\usepackage{mathrsfs}%
\usepackage[title]{appendix}%
\usepackage{xcolor}%
\usepackage{textcomp}%
\usepackage{manyfoot}%
\usepackage{booktabs}%
\usepackage{algorithm}%
\usepackage{algorithmicx}%
\usepackage{algpseudocode}%
\usepackage{listings}%

\theoremstyle{thmstyleone}%
\theoremstyle{thmstyletwo}%

\theoremstyle{thmstylethree}%

\begin{document}

\title[Look into your Heart - Prototypes for a Speculative Design Exploration of Personal Heart Rate Visualization]{Look into your Heart - Prototypes for a Speculative Design Exploration of Personal Heart Rate Visualization}


\author[]{\fnm{Swaroop} \sur{Panda}}



\affil[]{\orgdiv{School of Computer Science}, \orgname{Northumbria University}}




\abstract{Personal heart rate data from wearable devices contains rich information, yet current visualizations primarily focus on simple metrics, leaving complex temporal patterns largely unexplored. We present a speculative exploration of personal heart rate visualization possibilities through five prototype approaches derived from established visualization literature: pattern/variability heatmaps, recurrence plots, spectrograms, T-SNE, and Poincaré plots. Using physiologically-informed synthetic datasets generated through large language models, we systematically explore how different visualization strategies might reveal distinct aspects of heart rate patterns across temporal scales and analytical complexity. We evaluate these prototypes using established visualization assessment scales from multiple literacy perspectives, then conduct reflective analysis on both the evaluation and the design of the prototypes. Our iterative process reveals recurring design tensions in visualizing complex physiological data. This work offers a speculative map of the personal heart rate visualization design space, providing insights into making heart rate data more visually accessible and meaningful.}

\keywords{Personal Data Visualization, Heart Rate Visualization, Speculative Design, Prototypes}



\maketitle

\section{Introduction}

A fitness tracker may record a heart rate spike to 120 BPM during a stressful afternoon meeting, yet the user viewing this data lacks the interpretive framework to understand what these patterns reveal about cardiovascular health, stress response, or autonomic nervous system function. Despite unprecedented access to personal heart rate data through ubiquitous wearable devices, most people cannot interpret the rich physiological narratives embedded within their own cardiovascular streams \cite{choe2014understanding, huang2015personal}. The continuous monitoring capabilities of modern devices generate vast amounts of temporal data, yet users remain largely unable to extract meaningful insights from these complex patterns. While modern smartwatches effortlessly capture beats per minute, heart rate variability, and recovery metrics, these numerical presentations fail to transform complex temporal patterns into actionable health insights that could inform daily wellness decisions \cite{perin2015personal, bent2020investigating,khundaqji2020smart}. The gap between data availability and interpretability continues to widen as devices become more sophisticated, creating a paradox where more data leads to less understanding.

This interpretation gap becomes particularly problematic because heart rate data contains extraordinary information about autonomic nervous system balance, circadian rhythm regulation, exercise adaptation, and emotional state fluctuations - insights that remain locked behind clinical terminology and statistical abstractions \cite{shaffer2017overview, Dong16092024}. These physiological signals encode stories about stress resilience, recovery capacity, and cardiovascular adaptation that could profoundly inform personal health decisions if made accessible through appropriate visual representations. Current consumer interfaces predominantly rely on basic line charts and numerical dashboards that inadequately represent the multiscale temporal dynamics of cardiovascular function, leaving users disconnected from understanding their own biological rhythms despite carrying sophisticated physiological monitoring systems daily \cite{li2023heart,de2024state}. The visualization approaches employed by most consumer devices prioritize simplicity over insight, reducing complex physiological phenomena to oversimplified metrics that obscure rather than reveal the underlying patterns of cardiovascular function\cite{rodrigues2022hrv,stone2021assessing,kim2018stress}.

Speculative methods have become increasingly vital in HCI, enabling researchers to explore possible futures and challenge existing paradigms beyond immediate technological constraints. Speculative design fiction allows researchers to create diegetic prototypes that embody potential futures \cite{bleecker2022design,tanenbaum2012design}, while critical design questions assumptions about technology use \cite{dunne2013speculative}). These methods have proven particularly valuable for exploring ethical implications of emerging technologies \cite{wong2018speculative}, understanding long-term consequences of design decisions \cite{odom2016research}, and engaging stakeholders in conversations about preferable futures\cite{elsden2017speculative}.LLMs have emerged as powerful tools for speculative design, enabling researchers to generate alternative scenarios and narratives at scale \cite{blythe2025artificial} and create interactive design fictions \cite{panda2024thought} that respond dynamically to user input. These LLMs facilitate rapid prototyping of speculative concepts by generating diverse perspectives, potential user dialogues, and future-oriented design provocations that help designers explore the space of possibilities more comprehensively.

We present a speculative design exploration of personal heart rate visualization through five prototypes, each revealing different heart rate patterns across temporal scales and analytical dimensions. Using speculation as a framework allowed us to explore visualization possibilities while avoiding the ethical and practical constraints of traditional physiological data collection. Working with physiologically-informed large language model generated datasets, we developed prototypes spanning multiple analytical perspectives: heatmaps for circadian rhythms, recurrence plots for temporal patterns, spectrograms for frequency analysis, T-SNE for pattern relationships, and Poincaré plots for variability assessment. We evaluated these prototypes using established visualization scales through four user personas representing different visualization literacy levels. Our analysis revealed preliminary patterns that begin to map the design space for personal heart rate visualization, highlighting key tensions and opportunities in translating complex heart rate data into accessible visual forms.

The contributions of this paper include:
\begin{enumerate}
\item Five systematic prototypes exploring distinct approaches to personal heart rate visualization across multiple temporal scales and analytical frameworks
\item A speculative large language model-driven evaluation methodology for these prototypes employing established visualization assessment scales and visualization literacy tests
\item A speculative design space and preliminary design implications for personal heart rate visualization that emerges from the prototype design \& analysis
\end{enumerate}

\section{Background}

\subsection{Personal Data Visualization}\label{personal}
Personal data visualization has emerged as a pivotal research domain addressing the unique challenges of designing visualizations for individual reflection and self-understanding \cite{huang2015personal,pousman2007casual}. Aseniero et al's work demonstrates how Activity River enables individuals to plan, log, and reflect on self-defined activities through temporal visualization, supporting dynamic and continuous reflection patterns including forward shifting, backward shifting, and activity replacement \cite{aseniero2020activity}. Perin's research on personal data physicalization reveals that students learn about data visualization, design creatively, and gain self-knowledge by representing personal data in physical forms, providing evidence that tangible representations benefit even technically-savvy audiences \cite{perin2021students}. The Quantified Self movement has significantly influenced this field, with researchers examining how people collect, explore, and gain insights from personal tracking data \cite{choe2014understanding,choe2015characterizing}.Studies of Quantified Self presentations identify eight distinct insight types including self-reflection, trend analysis, and correlation discovery, highlighting the need for visualization systems that support reflection, valid insight generation, and communication \cite{choe2015characterizing}. Casual information visualization principles have proven essential for personal contexts, emphasizing aesthetics, playfulness, and pleasure over analytical depth to engage broader audiences \cite{pousman2007casual}. Research demonstrates that personal visualization requires different design considerations than traditional information visualization, including support for flexible data collection, narrative formation, and emotional engagement \cite{huang2015personal}. Contemporary work explores patient-generated data visualizations, revealing the complexity of designing for heterogeneous stakeholders including patients, healthcare providers, and systems \cite{rajabiyazdi2020exploring}. The field has evolved from simple tracking interfaces to sophisticated tools supporting sensemaking, with studies showing that visual data exploration enables rich self-reflection and insight generation when properly scaffolded \cite{choe2017understanding}. 

This research trajectory establishes personal data visualization as a distinct subdomain requiring specialized approaches that prioritize individual meaning-making, reflection, and empowerment over traditional analytical objectives\cite{van2005value}. The basis for our research on personal data visualization is grounded in these approaches and findings.


\subsection{Heart Rate Data Visualization} \label{hrvdatareview}
Heart rate data analysis and visualization has emerged as a rapidly evolving interdisciplinary field, combining advances in signal processing, human-computer interaction, and clinical informatics to enable comprehensive cardiovascular monitoring across consumer and medical contexts\cite{xue2022understanding,holzinger2013interactive,guo2020heat}. Recent research demonstrates significant technical advances in visualization techniques, with innovative approaches including circular ECG projections (Star-ECG visualization) \cite{9630507}, cohort-based visual analytics frameworks (GUCCI) \citep{meuschke2021gucci}, and immersive VR environments for cardiac surgical planning \citep{sadeghi2020immersive}, representing a shift from traditional waveform displays toward more intuitive and interactive representations. Signal processing methodologies have evolved substantially, incorporating deep learning architectures such as convolutional neural networks and LSTM models achieving over 98\% accuracy for arrhythmia detection \citep{oh2018automated}, while edge computing implementations enable real-time processing with sub-millisecond latency on resource-constrained wearable devices. Heart rate variability analysis has expanded beyond traditional time and frequency domain measures to include nonlinear methods like Poincaré plot analysis, detrended fluctuation analysis, and entropy measures, providing richer insights into cardiac autonomic function \citep{henriques2020nonlinear}. 

Within HCI, there is some (though limited) adoption of user-centered design principles in heart rate monitoring and visualization interfaces \citep{hassib2017heartchat}, with studies highlighting the need for personalized, accessible, and privacy-preserving visualization systems that support both clinical decision-making and patient engagement \citep{tadas2020barriers,rook2025heart}. Wearable device integration has achieved clinical-grade accuracy in consumer devices, with systems like Apple Watch obtaining FDA clearance for atrial fibrillation detection, though challenges remain regarding battery optimization, multi-modal sensor fusion, and privacy preservation \citep{shahid2025diagnostic}. 

Emerging trends include the integration of explainable AI for interpretable cardiac monitoring \cite{elvas2025role}. Federated learning approaches are applied for privacy-preserving clinical research\cite{teo2024federated}. These trends point toward an ongoing convergence of visualization, machine learning, and ubiquitous computing technologies within cardiovascular health monitoring, thereby underscoring the critical importance of this research in personal health data visualization.


\subsection{Prototypes \& Speculative Design exploration in HCI}


Prototypes and speculative design explorations constitute essential methodological contributions in HCI research, offering tangible pathways for investigating complex socio-technical futures and possibilities that cannot be adequately explored through traditional empirical methods alone. Zimmerman et al. \cite{zimmerman2007research} established Research through Design (RtD) as a rigorous methodology where artifacts embody theoretical insights and technical possibilities, transforming abstract concepts into concrete objects that stakeholders can experience, critique, and build upon. This approach has proven particularly valuable when addressing "wicked problems" in HCI - those ill-defined challenges lacking clear solutions where prototypes serve as material hypotheses about preferable futures \cite{buchanan1992wicked}. The legitimacy of prototype contributions extends beyond mere technical demonstrations. Höök and Löwgren \cite{hook2012strong} argue for ``strong concepts" as intermediate-level knowledge residing between specific instances and abstract theories, with prototypes serving as vehicles for articulating these transferable design insights. Similarly, Odom et al. \cite{odom2016research} demonstrate how research products can generate knowledge through longitudinal deployment, revealing insights about technology adoption and appropriation that emerge only through sustained use. The RtD paradigm emphasizes that prototypes function simultaneously as inquiry instruments for exploring under-constrained problems and as exemplars providing concrete instantiations that facilitate knowledge transfer \cite{gaver2012annotated, koskinen2011design}. 

Speculative and critical design approaches further expand the epistemic value of prototypes by deliberately challenging assumptions about technology's role in society. Dunne and Raby \cite{dunne2024speculative} position speculative design as creating design fictions that provoke discussion about alternative technological futures, while Auger \cite{auger2013speculative} argues these approaches make complex issues tangible and debatable through material speculation. Wong and Khovanskaya \cite{wong2018speculative} demonstrate how speculative design in HCI enables exploration of ethical implications and value tensions before technologies become entrenched in society. Pierce et al. \cite{pierce2015expanding} further articulate how design fiction and speculative design serve as "research through provocative design" that generates knowledge by challenging existing paradigms. Contemporary HCI increasingly recognizes that prototype contributions advance the field when they embody novel interaction techniques, demonstrate improved functionality, reveal behavioral insights, or articulate design spaces through systematic exploration \cite{fallman2003design}. Greenberg and Buxton \cite{greenberg2008usability} argue that premature demands for usability validation can stifle innovation, advocating instead for recognizing prototypes' value in opening new design territories. The evaluation criteria for prototype contributions have evolved to encompass not only functional performance but also their capacity to reframe problems, inspire further research, and transfer insights across domains \cite{bardzell2012critical, gaver2012should}.


This methodological pluralism strengthens HCI's capacity to address emerging challenges in personal health informatics, where traditional evaluation methods may be ethically problematic or practically infeasible. Prototypes and speculative explorations enable researchers to investigate sensitive domains, map design spaces, and articulate possibilities that inform both research and practice communities - contributions that complement but cannot be replaced by empirical studies alone. We use such prototypes for speculative explorations in heart rate data visualization.

\section{Method}


\subsection{Using LLMs for Speculative Design Exploration}

LLMs serve as valuable instruments for speculative design exploration in HCI, particularly when investigating sensitive health data domains where traditional collection methods face ethical, practical and often economic constraints \cite{hamalainen2023evaluating}. LLMs enable generation of physiologically plausible synthetic datasets that preserve temporal patterns and statistical properties essential for prototype development, while their code generation capabilities accelerate iterative design refinement \cite{kim2024health}. For our work, LLMs generate the synthetic data by writing code.

We also use LLMs to evaluate our prototypes. Recent work demonstrates LLMs' effectiveness in simulating diverse user perspectives through persona-based evaluation \cite{schuller2024generating,panda2024llms, hamalainen2023evaluating}, providing insights into visualization interpretation across varying expertise levels without recruiting participants . This approach aligns with speculative design methodologies in HCI that prioritize mapping possibility spaces over immediate feasibility \cite{elsden2017speculative, wong2018speculative}, enabling researchers to identify design tensions and explore alternative futures while maintaining rigor through structured evaluation frameworks. 

By positioning LLMs as design partners in the research process, we can systematically investigate visualization approaches that would otherwise remain unexplored due to privacy concerns or economic/resource limitations \cite{pan2025agentcoord}.

The LLM used in this research is Claude \citep{anthropic2025claude} Sonnet 4.

\subsubsection{Available Data Collection Methods}
The visualization prototypes presented in this work utilize data types that can be readily captured by commercially available cardiovascular monitoring devices. Contemporary consumer wearables, including chest strap monitors and smartwatches, provide continuous heart rate measurements with sufficient temporal resolution for pattern analysis. Clinical-grade equipment such as ECG monitors capture detailed cardiac electrical activity, enabling heart rate variability analysis and arrhythmia detection. Pulse oximeters offer non-invasive continuous monitoring suitable for long-term data collection. 

While our prototypes employ synthetic data for (privacy preserving) exploration, the underlying data structures and temporal characteristics align with outputs from these existing technologies, ensuring that our visualization approaches remain applicable to real-world physiological monitoring scenarios.
\subsection{Development of the Prototypes}


The development of these prototypes followed a systematic workflow applied consistently across all five visualization approaches. We began by establishing the user rationale for each visualization (why users might find value in viewing this aspect/analysis of heart rate data), identifying specific analytical needs such as understanding heart rate variability patterns or detecting temporal anomalies that each approach might address. This initial assessment ensured that each prototype served some purpose in revealing distinct aspects of heart rate data rather than merely demonstrating technical capability.

Following this conceptual grounding, we generated physiologically-informed synthetic datasets using LLMs through carefully crafted prompts that specified relevant cardiovascular parameters and temporal characteristics. Each dataset was exported in CSV format and processed through Python scripts that handled data preprocessing, parameter configuration, and transformation into the appropriate structure for each visualization technique.

The implementation leveraged established Python visualization libraries, primarily Matplotlib and Seaborn, to transform the heterogeneous temporal data into distinct visual representations. While specific implementation details varied across prototypes the overall development pipeline remained consistent. 

The prototype development process tried to balanced scientific accuracy with aesthetic appeal, recognizing that personal health visualizations serve a fundamentally different purpose than clinical displays (Section \ref{personal}). Since these visualizations are designed for individual consumption rather than medical practitioner analysis, we incorporated artistic considerations (directly in the python scripts) to some prototypes alongside technical requirements. This approach involved drawing inspiration from existing art forms and visual design traditions to create representations that users might find visually engaging and emotionally resonant. 

The integration of artistic considerations into the prototype was guided by the goal of increasing engagement rather than adherence to a specific theoretical framework \cite{lan2025more}. We sought to develop visualizations that individuals would willingly interact with regularly - perhaps even display as ambient information in personal spaces - while maintaining the data integrity necessary for meaningful health insights. This dual focus on aesthetic appeal and analytical utility reflects our understanding that effective personal health visualization must engage users emotionally and intellectually\cite{turchioe2019systematic}, transforming complex physiological data into compelling visual narratives that invite exploration rather than intimidate through clinical formality.

\subsection{Evaluation of the Prototypes}
We evaluate the visualizations through a two-pronged approach using LLMs to simulate user assessments. We instructed LLMs to adopt four distinct user personas \cite{chang2008personas,schuller2024generating} representing different levels of visualization literacy, from novice to expert users. We first provided comprehensive context about visualization literacy frameworks to ensure the LLMs could authentically embody each persona's perspective and capabilities. These simulated users then evaluated each prototype using three established scales: BeauVis for aesthetic quality and PreVis for perceptual effectiveness. This multi-perspective evaluation approach allowed us to systematically explore (speculate) how users with varying expertise levels might perceive and interpret each visualization technique, providing insights into accessibility and effectiveness across diverse user populations.

We employ a persona-based evaluation approach with LLMs for two primary reasons. First, rather than examining how LLMs themselves assess these visualizations, our interest lies in speculating human evaluation. Second, we seek to explore how individuals with varying levels of visualization literacy might interpret and assess these data representations. By instructing LLMs to adopt distinct user personas before evaluating the visualizations, we can simulate diverse human perspectives and literacy levels, making this methodology particularly well - suited to our research objectives.

\subsubsection{BeauVis Scale}
The BeauVis Scale represents a validated instrument developed by He et al.~\cite{he2023beauvis} for measuring aesthetic pleasure in data visualizations. This scale emerged from the critical need to address the lack of standardized aesthetic assessment tools in visualization research, where researchers previously relied on ad-hoc terminology or general aesthetic scales not specifically designed for visualizations. The original validated scale consists of five items measuring responses to "enjoyable," "likable," "pleasing," "nice," and "appealing" on 7-point Likert scales.  


\subsubsection{PreVis Scale}

The PreVis (Preference for Visualization) scale measures users' subjective preferences and satisfaction with data visualizations \cite{cabouat2024previs}. This six-item instrument evaluates multiple dimensions of visualization effectiveness, including suitability for the given task, appropriateness for the data type, and overall user satisfaction. Participants rate each statement using a Likert-type scale, providing insights into their perceived utility and willingness to recommend the visualization to others. The scale serves as a comprehensive measure of visualization preference, capturing both functional and experiential aspects of the user's interaction with the visual representation.


\subsection{Speculative Design Exploration}

The speculative design space emerged from systematic observations during the prototype design, development and evaluation and through identifying recurring design considerations across all five visualization approaches. We characterize this work as ``speculative" because our findings represent exploratory insights rather than empirically validated conclusions - the prototypes provide suggestive evidence for design possibilities rather than definitive proof of effectiveness (in contrast to \textit{validated} validations \cite{elmqvist2012patterns}). So our design implications and considerations remain deliberately preliminary, grounded in the specific prototypes we developed and the design principles they embody rather than comprehensive user studies or clinical validation. This speculative stance allows us to map a broader landscape of possibilities while acknowledging the inferential nature of our observations. 


\section{Prototypes}

The designation of this work as visualization ``prototypes" reflects both its current developmental stage and intended research contribution. These prototypes represent proof-of-concept implementations that demonstrate technical feasibility and design principles while acknowledging the need for subsequent validation and refinement. 

To restate, we make no claims regarding clinical effectiveness, as these visualizations have not undergone clinical validation by practitioners or formal medical frameworks. In optimal circumstances, medical-grade visualizations would be seamlessly integrated within electronic health records or clinical decision support systems, functioning as components of larger diagnostic ecosystems with appropriate regulatory compliance and healthcare data standards. However, our standalone implementations serve as essential intermediate steps, allowing for speculative exploration and refinement of visualization concepts before addressing system integration complexities.

The Institutional Ethics Committee approved the data generation procedure employed in this study. 

The prompts that were used to generate data were, 

\begin{quote}
    For \textbf{4.1} : \texttt{Generate a realistic week long dataset of heart rate measurements of a healthy human and with realistic daily patterns. CSV format with time in seconds and minutes, heart rate in bpm, timestamp columns.} \\
    For \textbf{4.1-4.5} : \texttt{Generate 7 days of realistic RR interval data with circadian rhythms, respiratory sinus arrhythmia, and random exercise periods. Include timestamp, rr interval ms, heart rate in bpm columns.}
\end{quote}

\subsection{Circadian Rhythm Heatmap}
\begin{figure}[!htb]
    \centering
    \framebox{
    \includegraphics[width=0.95\linewidth]{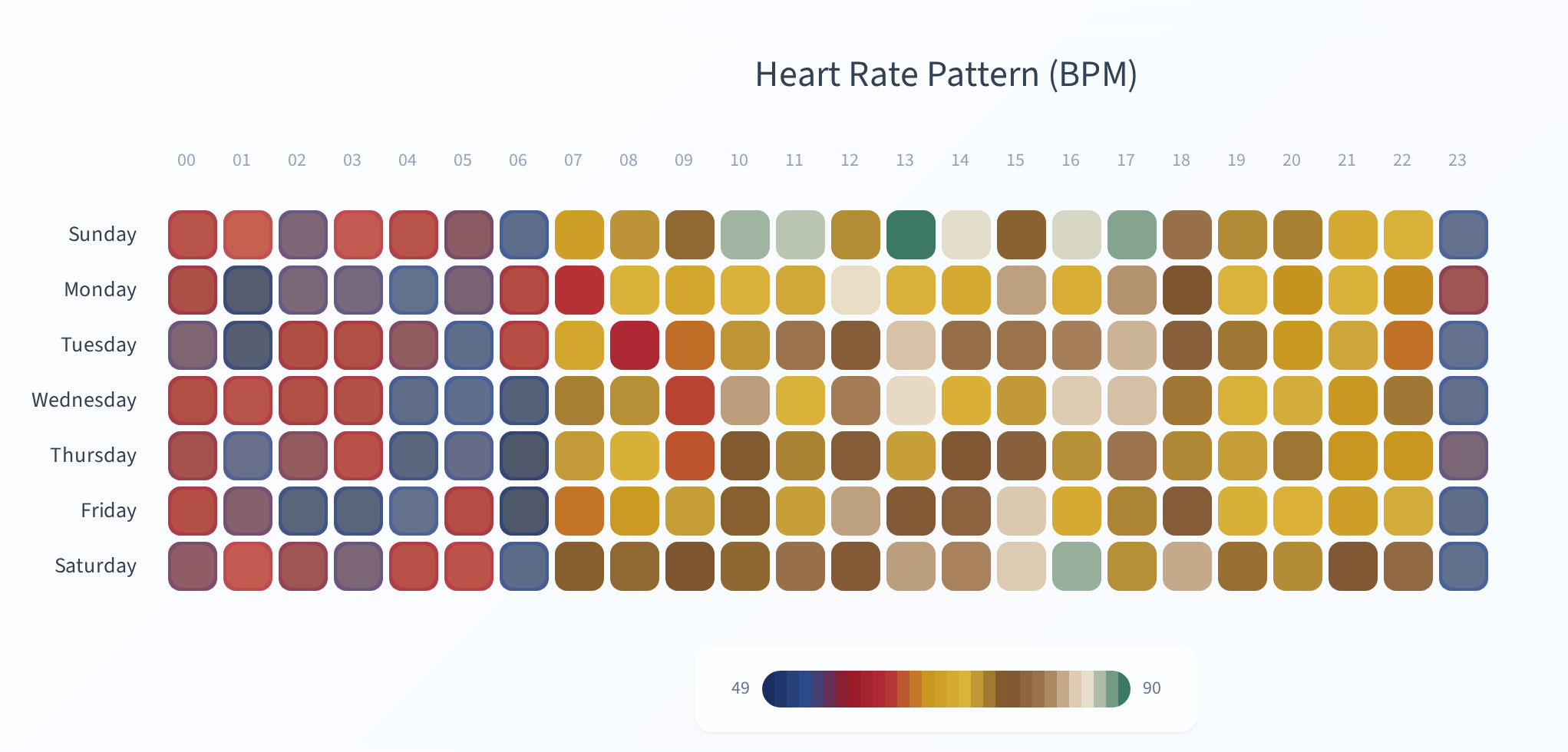}}
    \caption{Heart Rate in BPM for a Week}
    \label{fig:bpmweek}
\end{figure}

\begin{figure}[!htb]
    \centering
    \framebox{
    \includegraphics[width=0.95\linewidth]{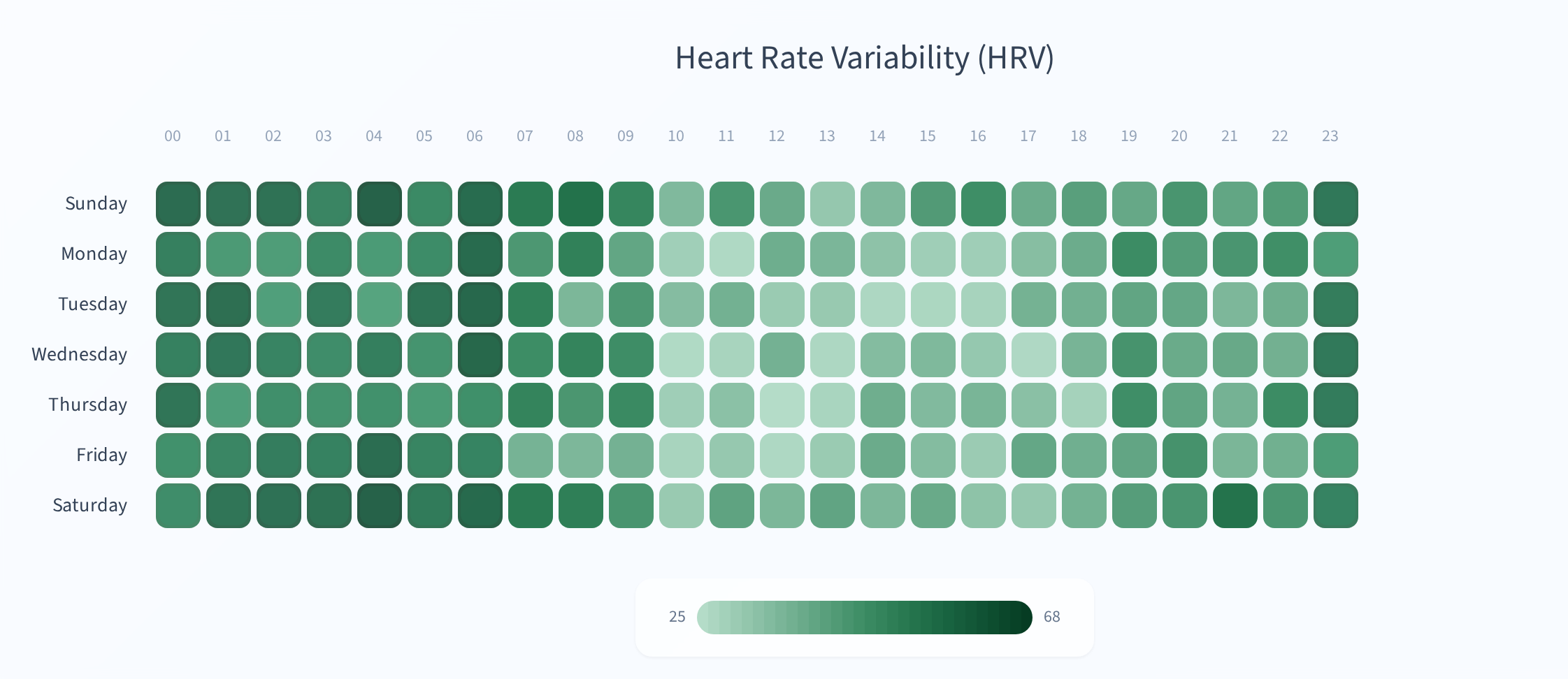}}
    \caption{Heart Rate Variability through the Week}
    \label{fig:hrvweek}
\end{figure}

The circadian rhythm heatmap prototype transforms weekly heart rate data into an intuitive temporal visualization that reveals daily physiological patterns through careful visual design. The implementation employs a sophisticated dual-heatmap approach: a primary matrix \ref{fig:bpmweek} displaying average heart rate patterns using a twilight-inspired color palette, and a secondary Heart Rate Variability (HRV) heatmap \ref{fig:hrvweek} using aurora-inspired greens to indicate heart rate consistency throughout the day. HRV measures the variation in time intervals between consecutive heartbeats, serving as a key indicator of autonomic nervous system balance and overall physiological resilience . Unlike simple heart rate measurements that average beats per minute, HRV captures the subtle millisecond-level fluctuations that reflect the dynamic interplay between sympathetic (stress) and parasympathetic (recovery) nervous system activity \cite{shaffer2017overview}.

The visualization's strength lies in its multi-layered information architecture. The main heatmap provides immediate pattern recognition users can instantly identify their sleep wake cycles, morning heart rate rises, and evening wind-down periods. Time zone indicators and sleep period overlays if added to these charts, can add contextual understanding without cluttering the primary data display. 

The square grid heatmaps provide yet another perspective, using color intensity to represent both heart rate magnitude and variability across hourly blocks. This format excels at revealing weekly patterns - users can quickly identify which hours consistently show elevated heart rates or increased variability. The discrete square format creates a more structured, calendar-like view that may feel more familiar to users accustomed to digital interfaces.

\begin{figure*}[!htb]
    \centering
    \framebox{
    \includegraphics[width=0.95\linewidth]{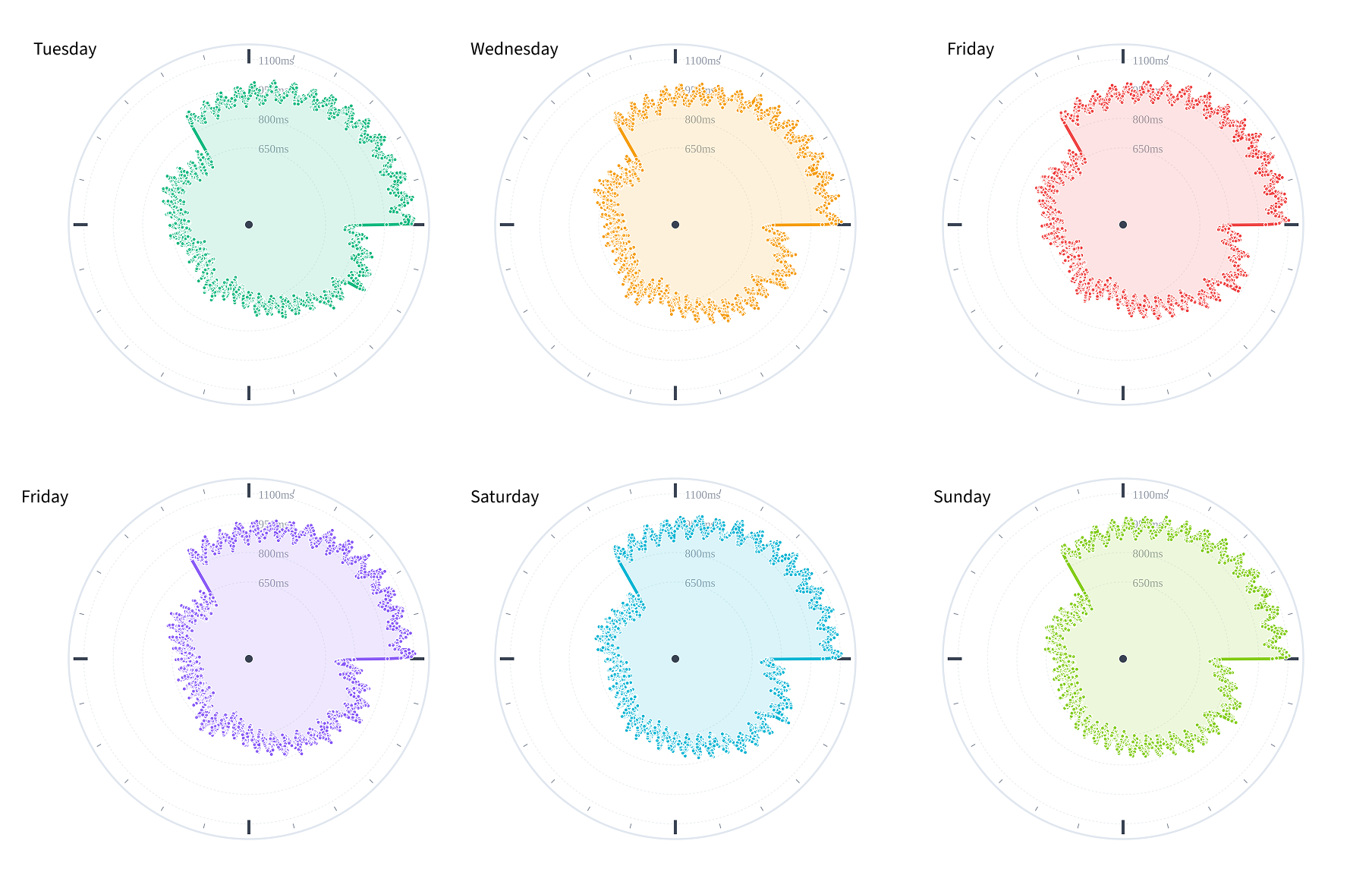}}
    \caption{Heart Rate Clock from Tuesday to Sunday}
    \label{fig:heartrateclocks}
\end{figure*}
\begin{figure}[!htb]
    \centering
    \framebox{
    \includegraphics[width=0.80\linewidth]{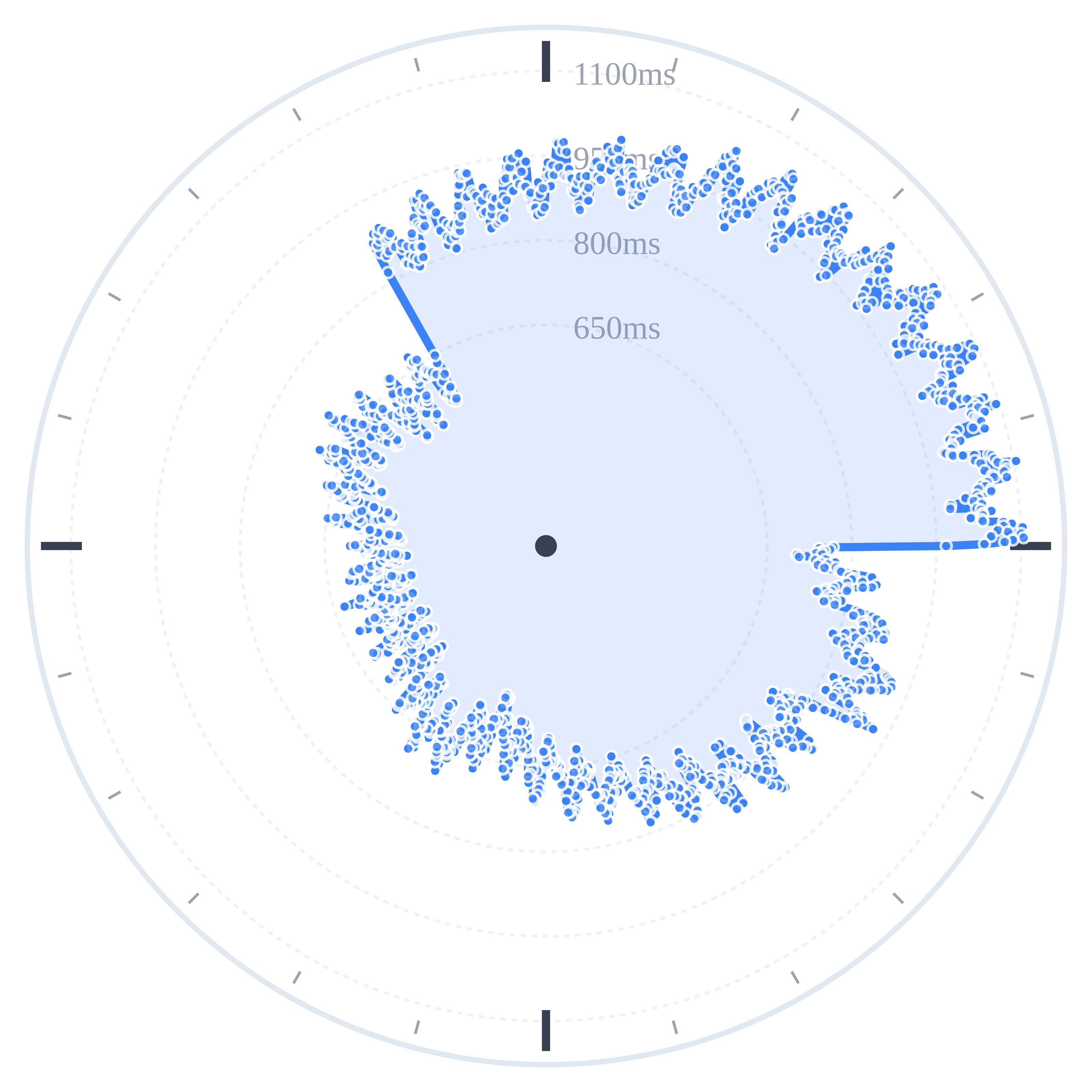}}
    \caption{Zoomed in Monday Heart Clock}
    \label{fig:zoomedheartrateclocks}
\end{figure}

The prototype also explores alternative representations through polar coordinate systems, as shown in the clock-face visualizations \ref{fig:heartrateclocks}. These radial plots map heart rate intervals onto a 24-hour clock face, with each point representing consecutive heartbeat intervals \ref{fig:zoomedheartrateclocks}. The resulting patterns create distinctive fingerprints for each day, with tighter clusters indicating more stable heart rate and scattered patterns suggesting greater variability. The use of different colors for each day enables quick visual comparison of daily cardiovascular patterns.

\subsection{Recurrence Plot}
\begin{figure}
    \centering
    \includegraphics[width=0.95\linewidth]{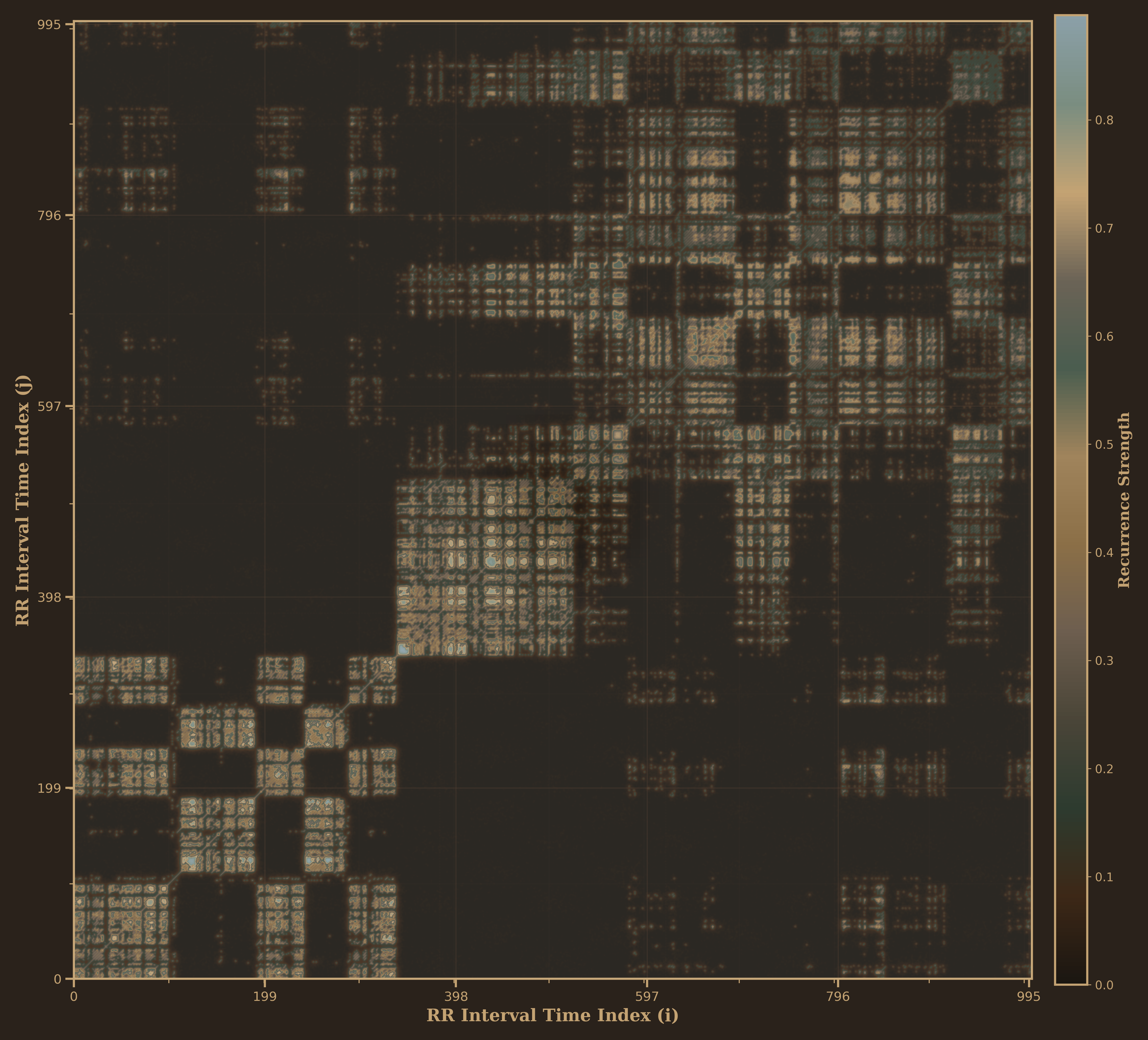}
    \caption{The \textit{Renaissance} Recurrence Plot. Both the axes are time indices.}
    \label{fig:recurrence-plot}
\end{figure}


Recurrence analysis provides a powerful non-linear method for examining temporal patterns in physiological time series data, particularly HRV measured through RR intervals - the time between consecutive heartbeats \cite{marwan2007recurrence}. In clinical contexts, RR interval data collected over extended periods (such as 7 days) contains rich information about autonomic nervous system function, cardiac health, and circadian rhythms \cite{malik1996heart, thayer2012heart}.

The recurrence plot visualization technique transforms one-dimensional time series data into a two-dimensional representation by examining when the cardiac system returns to previously visited states in phase space \cite{eckmann1987recurrence}. Each point $(i,j)$ in the recurrence matrix (Figure~\ref{fig:recurrence-plot} ) represents whether the cardiac state at time $i$ recurs at time $j$, based on a proximity threshold in the reconstructed phase space. Dark regions indicate recurring patterns, while light areas suggest unique or transitional cardiac states.

Figure~\ref{fig:recurrence-plot} presents a Renaissance-inspired artistic interpretation of RR interval recurrence analysis. The visualization employs Leonardo da Vinci's sfumato and chiaroscuro techniques to enhance pattern recognition through gradual tonal transitions and dramatic light-dark contrasts \cite{kemp2006leonardo}. The Mona Lisa color palette - derived from art historical analysis - provides intuitive visual encoding where darker tones represent recurring cardiac patterns and lighter regions indicate novel physiological states.

This recurrence plot displays a complex pattern with distinct structural features throughout the matrix. The plot shows concentrated regions of high recurrence activity (bright areas) interspersed with sparse zones (dark areas), creating a heterogeneous texture across the entire visualization. There are clear geometric patterns including rectangular blocks, diagonal lines, and clustered formations that vary in density and distribution. The overall structure suggests a system with mixed dynamics - some areas display organized, repetitive patterns while others appear more fragmented or irregular. The gradient coloring from dark to bright indicates varying degrees of recurrence strength, with the brightest regions representing the strongest pattern repetitions in the underlying time series.

Users can see their heart's hidden patterns through these visual plots, helping them understand their cardiovascular health in a way that goes beyond simple heart rate numbers. The patterns show how well their heart adapts to different situations throughout the day/week, making it easier to spot when something might be off with their heart rhythm regulation. \textit{Patients} can track their own progress during treatment or recovery by watching how their unique pattern changes over time, giving them a clearer picture of whether their heart health is improving or needs attention. These plots also help users better understand their personal risk levels and motivate them to take an active role in managing their heart health.




\subsection{Power Spectral Density Spectrogram}

\begin{figure*}
    \centering
    \includegraphics[width=0.95\linewidth]{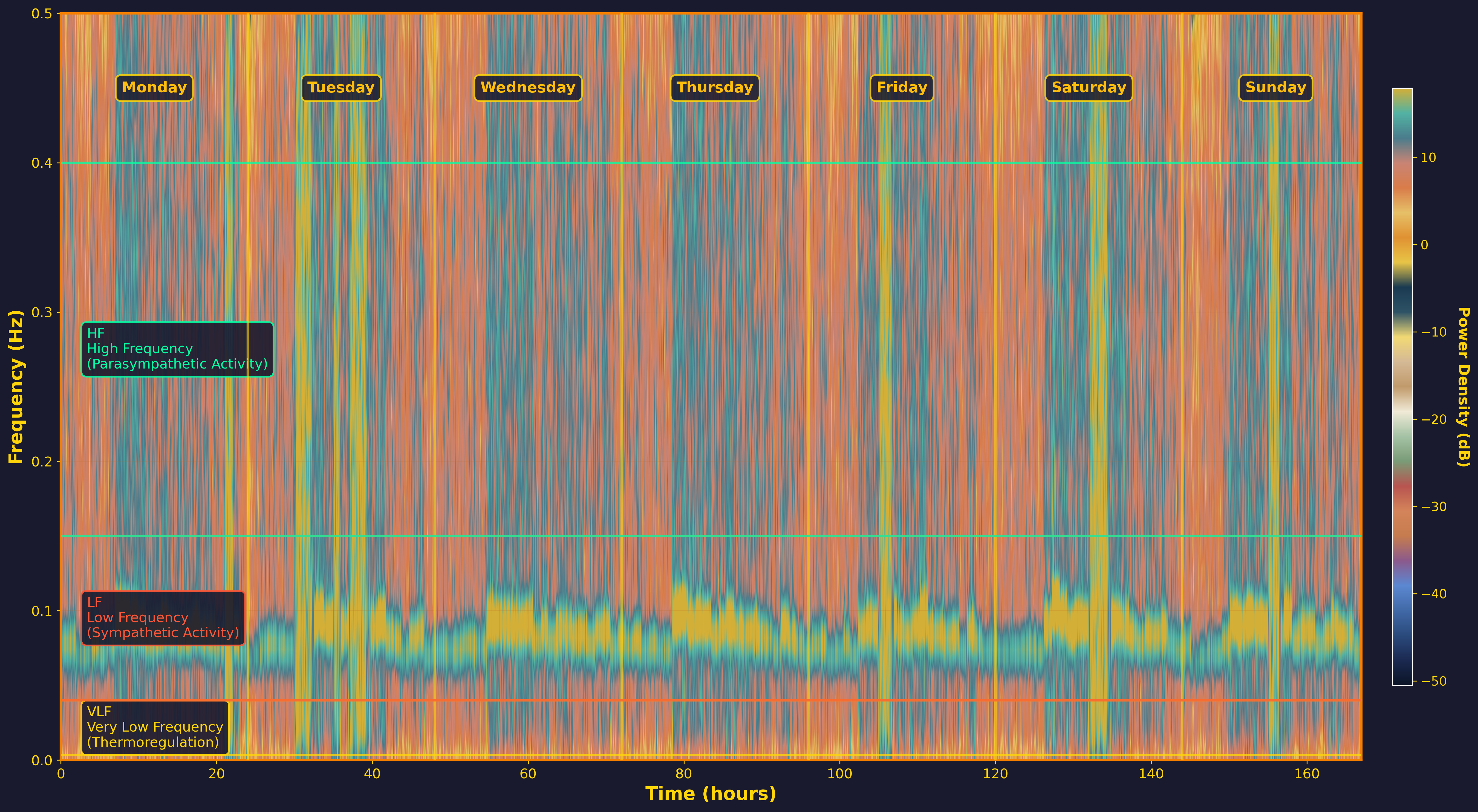}
    \caption{The \textit{Van Gogh} Spectrogram}
    \label{fig:spectrogram}
\end{figure*}

Spectrograms provide a fundamental time-frequency representation that reveals how the frequency content of physiological signals evolves over time \cite{cohen2014analyzing}. Unlike traditional frequency analysis that assumes stationarity, spectrograms employ sliding window techniques to capture dynamic changes in cardiac autonomic regulation across circadian cycles, sleep-wake transitions, and daily activities \cite{clifford2006advanced}.

Power Spectral Density (PSD) analysis decomposes heart rate variability into distinct frequency bands that correspond to specific physiological mechanisms. The Very Low Frequency (VLF: 0.003-0.04 Hz) band reflects thermoregulatory and hormonal influences, the Low Frequency (LF: 0.04-0.15 Hz) band primarily captures sympathetic nervous system activity, while the High Frequency (HF: 0.15-0.4 Hz) band represents parasympathetic (vagal) modulation associated with respiratory sinus arrhythmia \cite{shaffer2017overview}.

Figure~\ref{fig:spectrogram} presents a Van Gogh-inspired spectrogram visualization spanning seven days of heart rate monitoring. The artistic rendering employs the painter's characteristic vibrant palette and emotional intensity to transform clinical frequency analysis into an engaging visual narrative \cite{van2005van}. The temporal progression reveals circadian patterns of autonomic balance, with distinct frequency band boundaries clearly delineated through color-coded annotations.

The HRV spectrogram provides critical insights by revealing autonomic nervous system patterns over time. Circadian rhythms (clear day/night cycles across all frequency bands) help detect sleep disorders and circadian disruption. The LF band shows sympathetic activity - higher power during wake hours indicates normal function, while persistently elevated levels suggest chronic stress, cardiovascular risk, or overtraining in athletes. The HF band reflects parasympathetic activity with respiratory peaks around 0.25 Hz; strong HF power indicates good cardiac fitness and sleep quality, while reduced HF suggests autonomic neuropathy (common in diabetes) or cardiovascular disease. The VLF band captures slower oscillations related to thermo-regulation and hormonal cycles, with reduced VLF predicting risk in cardiac patients. 



\subsection{TSNE plots}

\begin{figure}[!htb]
    \centering
    \framebox{
    \includegraphics[width=0.95\linewidth]{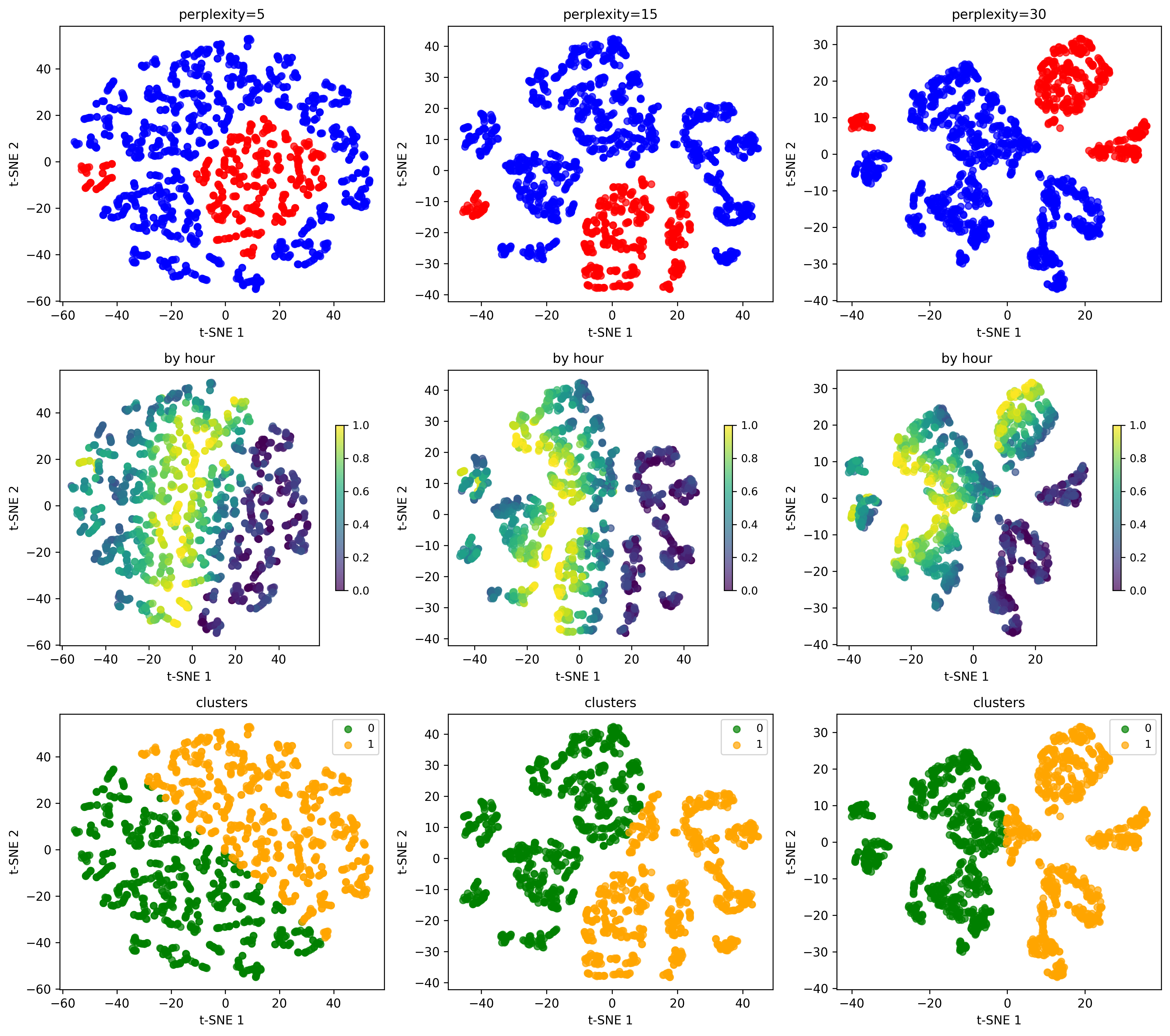}}
    \caption{TSNE Analysis. Different colors represent the different units visualized - weekdays, hours or days.}
    \label{fig:tsne2}
\end{figure}

Multivariate time series analysis of heart rate data involves examining multiple simultaneous physiological variables to uncover complex, non-linear relationships that univariate approaches cannot detect. When applied to extended RR interval recordings, this approach constructs feature vectors incorporating statistical measures (mean, variance, range), temporal dynamics (pNN50, RMSSD), and contextual information (circadian timing, weekend patterns) to capture the multidimensional nature of cardiovascular regulation \cite{brennan2001relationship}.

t-Distributed Stochastic Neighbor Embedding (t-SNE) represents a breakthrough non-linear dimensionality reduction technique that preserves local neighborhood structures while revealing global patterns in high-dimensional physiological data \cite{vandermaaten2008visualizing}. Unlike linear methods such as Principal Component Analysis, t-SNE can uncover complex, non-linear manifolds in cardiac feature spaces, making it particularly valuable for identifying subtle physiological states that may not be apparent through traditional analysis \cite{wattenberg2016use}.

The perplexity parameter in t-SNE controls the balance between preserving local versus global structure, with lower values emphasizing fine-grained clusters and higher values revealing broader organizational patterns \cite{wattenberg2016use}. Figure~\ref{fig:tsne2} demonstrates this principle across multiple perplexity settings (5, 15, 30), showing how different parameter choices reveal varying aspects of cardiac state organization. The color-coding by temporal features (hour of day, weekend status) reveals circadian clustering patterns and lifestyle-dependent physiological signatures.

\begin{figure}[!htb]
    \centering
    \framebox{
    \includegraphics[width=0.95\linewidth]{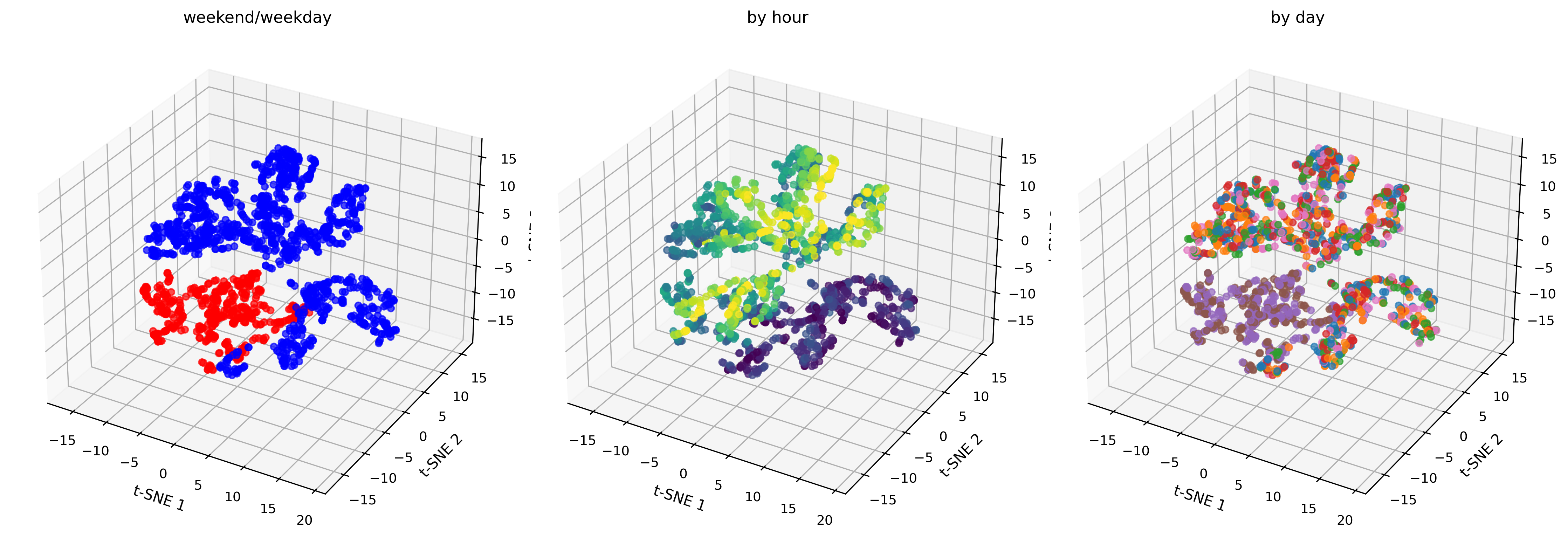}}
    \caption{TSNE in 3D. }
    \label{fig:tsne1}
\end{figure}

The three-dimensional t-SNE embedding (Figure~\ref{fig:tsne1}) provides additional structural insights, particularly valuable for understanding the complex topology of cardiac state spaces \cite{vandermaaten2008visualizing}. 

The clear separation between weekend and weekday patterns suggests that lifestyle factors create distinct physiological signatures detectable through multivariate HRV analysis. This capability has significant implications for personalized health monitoring, enabling the detection of stress patterns, sleep quality changes, and autonomic dysfunction through continuous cardiac surveillance.

\subsection{Poincaré Plot}

Poincaré plot analysis represents a geometric approach to heart rate variability assessment that visualizes the correlation structure between consecutive RR intervals through scatter plot representation \cite{brennan2001relationship, tulppo1996quantitative}. This technique plots each RR interval $(RR_n)$against the subsequent interval $(RR_{n+1})$, creating a two-dimensional phase space that reveals both short-term and long-term variability patterns in cardiac rhythm regulation \cite{kamen1996application}.

The geometric properties of the Poincaré plot provide clinically meaningful insights into autonomic nervous system function. The plot's shape, orientation, and dispersion patterns reflect the balance between sympathetic and parasympathetic influences on cardiac control \cite{brennan2001relationship}. Two primary quantitative measures characterize the plot: SD1 represents short-term variability (primarily parasympathetic activity) measured as the standard deviation perpendicular to the line of identity, while SD2 captures long-term variability (sympathetic-parasympathetic balance) as the standard deviation along the line of identity \cite{tulppo1996quantitative}.

\begin{figure}
    \centering
    \includegraphics[width=0.95\linewidth]{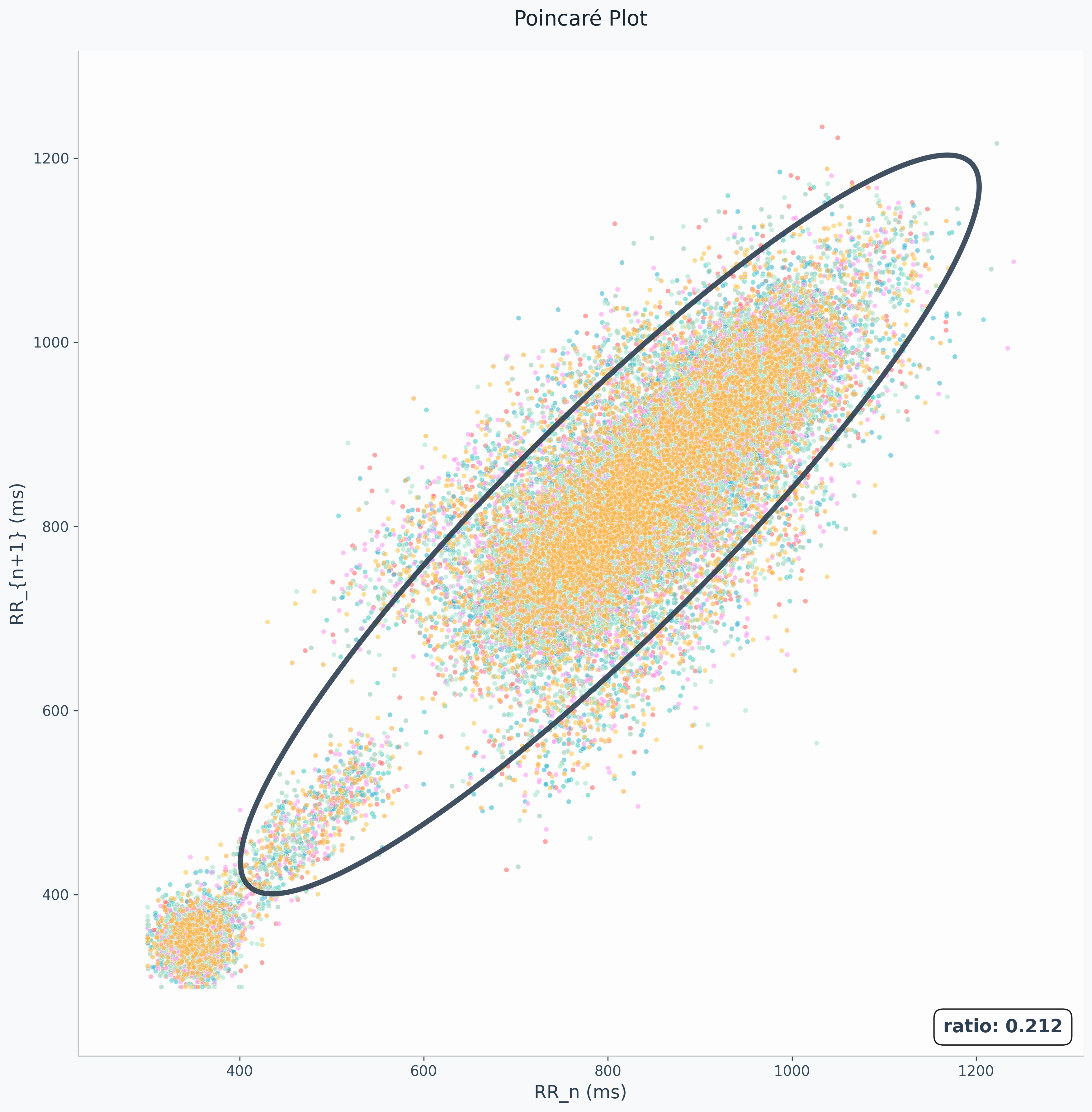}
    \caption{Poincaré plot analysis of consecutive RR intervals}
    \label{fig:poincare-plot}
\end{figure}

Figure~\ref{fig:poincare-plot} demonstrates a vibrant interpretation of Poincaré analysis applied to seven - day cardiac monitoring data. The visualization employs multiple color layers to represent data density while maintaining scientific accuracy through proper scaling and confidence ellipse fitting. The elliptical boundary represents the 95\% confidence region, indicating the normal range of cardiac variability for the monitored period.

The SD1/SD2 ratio provides a normalized measure of cardiac autonomic balance, with values typically ranging from 0.1 to 0.5 in healthy individuals \cite{brennan2001relationship}. This geometric approach offers several advantages for clinical assessment: it provides intuitive visual feedback about cardiac health status, enables rapid identification of pathological patterns, and facilitates patient education through accessible visual representation of complex physiological dynamics \cite{kamen1996application}. 

The technique proves particularly valuable for detecting early signs of autonomic dysfunction, monitoring treatment effectiveness, and assessing cardiovascular risk in both clinical and remote monitoring contexts.

\section{Evaluation}

We evaluated the visualization prototypes using two established assessment scales: BeauVis and PreVIS. Our evaluation methodology employed LLMs to simulate user responses across varying levels of visualization literacy, providing the LLM with the visualization images for analysis. To ensure systematic evaluation from multiple perspectives, we instructed the LLM to construct four distinct user personas based on established visualization literacy frameworks \cite{lee2016vlat,pandey2023mini}, with each persona representing a different expertise level ranging from novice to expert users. We then evaluate the visualizations across the two scales using each of the generated personas.  


The prompts used to evaluate are,

\begin{quote}
    \texttt{Generate four distinct user personas who consume heart rate data, differentiating them by their levels of visualization literacy using the Mini-VLAT framework as the reference. For each persona, provide a fictional name, demographic background, their specific visualization literacy level based on Mini-VLAT dimensions, and their goals and motivations for using heart rate data. Describe the typical context in which each persona engages with heart rate data (for example, personal health tracking, athletic training, or medical monitoring), along with the challenges or limitations they face when interpreting visualizations. Indicate their preferred visualization types or design features, if applicable. Present the four personas in a structured format.}
\end{quote}



\subsection{Results \& Analysis}

The PREVis functional assessment demonstrates a systematic relationship between visualization literacy levels and comprehension performance across all five heart rate visualization prototypes. High literacy users achieved mean scores of 5.3-7.0 on functional visualizations, while low literacy users scored consistently at 1.0, representing a 4-6 point performance gap. The Heatmap (\ref{fig:bpmweek}) and Poincaré Plot (\ref{fig:poincare-plot}) showed the highest accessibility (mean scores of 4.1 and 4.1 respectively), while the Recurrence Plot (\ref{fig:recurrence-plot}) and t-SNE plot (\ref{fig:tsne2}) created the steepest literacy-dependent performance gradients, with moderate-high literacy users scoring 2.5-3.0 points lower than experts on these technical visualizations. The Layout subscale (PREVis) consistently produced the lowest scores across all literacy levels and visualization types (range: 1.0-6.0), indicating universal challenges with visual complexity regardless of user expertise. Technical visualizations requiring specialized domain knowledge (Spectrogram, Recurrence Plot, t-SNE ) showed particularly steep performance drops between moderate-high and moderate literacy levels, suggesting critical comprehension thresholds.

BeauVis aesthetic ratings revealed parallel patterns with strong positive correlations to functional comprehension in PREVis. High literacy users averaged 4.96 points compared to 1.32 for low literacy users - a 3.64-point aesthetic appreciation gap. The Poincaré Plot achieved the highest cross-literacy aesthetic rating (4.0 average), followed by the Heart Rate Heatmap (3.5 average), while the Recurrence Plot (2.25 average) and t-SNE (2.1 average) received progressively lower ratings. Statistical analysis reveals strong positive correlations between PREVis functionality scores and BeauVis aesthetic ratings within individual visualizations, suggesting comprehension barriers directly impact aesthetic pleasure. Technical complexity created aesthetic penalties even for expert users, with advanced visualizations scoring 1-2 points lower than basic formats on aesthetic measures. The consistency of ratings across all five BeauVis dimensions (enjoyable, likable, pleasing, nice, appealing) indicates that aesthetic perception operates holistically rather than through independent evaluative criteria, with cognitive load probably affecting aesthetic perception independently of pure comprehension ability.


The combined PREVis and BeauVis findings reveal critical design imperatives for health data visualization systems. The strong correlation between functional comprehension and aesthetic appreciation (r=0.85-0.90) demonstrates that visualization literacy barriers create compounding engagement problems - users who cannot understand complex visualizations also find them unpleasant, reducing motivation for sustained health monitoring. This necessitates literacy-adaptive interface design where visualization complexity matches user capabilities. Simple formats like the Heart Rate Heatmap and Poincaré Plot maintain both functionality and appeal across user groups, while technical visualizations (Recurrence Plot, Spectrogram, t-SNE) should be reserved for expert users or implemented through progressive disclosure mechanisms. The universal Layout challenges indicate that successful personal health visualization must fundamentally rethink information architecture rather than adapting clinical tools for consumer contexts. 

These findings support tiered design approaches emphasizing clarity and reduced visual complexity to preserve both comprehension and aesthetic satisfaction across diverse user populations.

The Previs and BeauVIS scales are in Appendix A, the personas generated by the LLM are in appendix B, and the LLM responses to the scales are in appendices C and D.

\section{Preliminary Design Implications}

We developed these prototypes to establish a speculative design space for heart rate visualization, tackling the critical challenge of transforming physiological data into meaningful personal health insights. As stated earlier, the speculative dimension stems from integrating LLMs into our methodology - both for generating synthetic data and evaluating visualization effectiveness.

The design space we explore encompasses multiple interrelated dimensions. 

The first dimension involves navigating the fundamental tension between medical accuracy and personal meaning-making. Heart rate data represents serious physiological information requiring clinical precision, yet personal health visualization must also accommodate individual preferences, aesthetic sensibilities, and diverse interpretive frameworks. This echoes broader challenges in personal informatics where data must be both scientifically valid and personally resonant; as evident from the results. High literacy users achieved 5.3-7.0 PREVis scores on technical visualizations like the Spectrogram and Recurrence Plot, indicating successful clinical interpretation. However, moderate literacy users scored only 1.0-2.3 on these same visualizations, while low literacy users consistently scored 1.0, essentially failing to extract any meaningful information. This 4-6 point gap suggests that clinically accurate visualizations actively exclude most users from interpreting their own physiological data. The BeauVis findings reinforce this challenge: technical visualizations received poor aesthetic ratings even from expert users (HRV Spectrogram: 5.4 for experts vs. 1.0 for beginners; Recurrence Plot: 4.2 vs. 1.0), indicating that clinical accuracy often comes at significant aesthetic cost. The Poincaré Plot emerges as a promising compromise, maintaining the highest cross-literacy aesthetic appeal (4.0 average) while preserving clinical validity for heart rate variability analysis, suggesting that ``data portraits" \cite{lan2025more} are achievable when visual familiarity (scatter plot format) is combined with meaningful clinical content.


The second dimension addresses data integration and contextualization. Modern wearables capture heart rate alongside numerous other physiological and behavioral metrics - sleep patterns, physical activity, stress indicators, and environmental factors (Section \ref{hrvdatareview}). Effective personal health visualization must consider how heart rate data interconnects with these diverse streams to reveal holistic patterns about health and wellbeing. This multidimensional integration challenge requires careful consideration of visual encoding strategies, interaction techniques, and narrative structures that can coherently represent complex relationships without overwhelming users. The visualization must support both focused analysis of cardiovascular patterns and broader exploration of how HRV relate to daily activities, emotional states, and long-term health trajectories. The complexity penalty observed in the t-SNE Analysis directly addresses multidimensional integration challenges. This visualization, designed to show relationships between multiple physiological parameters, achieved the lowest aesthetic ratings (2.1 average) and caused comprehension failures even among moderate-high literacy users (2.7 PREVis scores vs. 5.7 for simpler visualizations). The Layout subscale consistently produced the lowest scores (1.0-6.0 range), indicating that current approaches to integrating multiple data streams overwhelm users regardless of expertise level. However, the Heatmap's relative success (3.5 BeauVis average, 4.1 PREVis average) demonstrates that temporal integration - showing heart rate patterns across weekly cycles - remains accessible while providing meaningful contextual information about daily activity patterns. 

The third dimension encompasses the technological and temporal aspects of visualization deployment. Personal health data visualization must adapt across multiple viewing contexts - from \textit{glanceable} smartwatch displays to detailed tablet interfaces to ambient home displays. This platform diversity demands responsive design strategies that maintain coherence while optimizing for each device's affordance and constraints. Furthermore, the temporal granularity of visualization updates presents critical design decisions: should visualizations emphasize real-time monitoring, daily summaries, weekly patterns, or longer-term trends? Different update frequencies serve distinct reflection and intervention purposes, from immediate biofeedback during exercise to monthly progress tracking for cardiovascular fitness improvement. Even expert users struggled with visual complexity, scoring 2.7-6.0 on Layout across different visualizations, suggesting that current dense, multi-panel approaches are unsuitable for smaller displays or \textit{glanceable} interfaces. 

The fourth dimension, perhaps most significantly, reconceptualizes the fundamental purpose and audience of medical data visualization. Traditional medical visualization assumes expert interpretation by healthcare professionals analyzing patient data within clinical contexts. In contrast, personal heart rate visualization inverts this relationship, positioning patients as primary interpreters of their own physiological data. This paradigm shift from clinician-centered to patient (or user) -centered visualization represents a fundamental re-imagining of medical data presentation. It requires developing new \textit{visual languages/patterns/grammars} that preserve medical validity while being accessible to non-experts, creating interpretive frameworks that empower rather than overwhelm users, and establishing trust mechanisms that help users distinguish between normal variations and concerning patterns. This patient-empowerment approach could potentially transform healthcare engagement, shifting from passive receipt of medical interpretation to active participation in health monitoring and decision-making. The challenge lies not merely in simplifying complex data but in creating visualization systems that support progressive skill development, allowing users to deepen their understanding of their cardiovascular patterns over time while maintaining appropriate boundaries around self-diagnosis and medical interpretation.  The 4-6 point PREVis performance gaps between expert and general users on technical visualizations (Recurrence Plot: 5.3 vs. 1.0; Spectrogram: 5.7 vs. 1.0) demonstrate that traditional medical visualization languages are fundamentally inaccessible to patients as primary interpreters.
The strong correlation between comprehension and aesthetic appreciation validates the patient (user) - empowerment approach: users cannot develop progressive skill with visualizations they find incomprehensible or unpleasant. The data reveals that moderate literacy users show the steepest aesthetic drops for technical content (Heart Rate Heatmap: 2.4 vs. Recurrence Plot: 1.2), suggesting that the transition from basic to advanced interpretation requires careful scaffolding rather than direct exposure to clinical-grade visualizations. 

\section{Discussion}
This work contributes artifacts to HCI research \citep{wobbrock2016research} and also aligns with Ringfort-Felner et al.'s speculative design quality taxonomy \cite{ringfort2025quality}. We demonstrate speculative quality through fictional explorations (data is LLM generated) of heart rate visualization. Our five prototypes exhibit discursive quality as working implementations with internal coherence and sufficient technical detail for evaluation. The physiologically-informed synthetic data and established HRV analysis methods demonstrate process quality, while LLM-based prototype evaluation represents a methodological innovation for design exploration. The work operates at the intersection of "speculative-experienceable-grounded" design \citep{ringfort2025quality}, balancing intellectual rigor with creative freedom while maintaining reflexivity about the nature of insights generated.
This research makes a unique contribution by combining speculative design methodology with quantitative analysis of LLM-generated data, including statistical evaluation. While we categorize this work as speculative because it lacks grounding in actual user data, we systematically employ LLMs in data generation and prototype evaluation. This approach might elicit ``AI shaming" concerns \cite{sarkar2025ai}. While AI-assisted research sometimes faces skepticism, we believe it opens doors to investigations that would otherwise be impossible. Though it can't replace primary measurements needed for clinical validation, it excels at speculative design exploration - mapping possibilities rather than proving clinical effectiveness. This transparency shows how AI tools, when properly acknowledged, enable rigorous research where traditional methods hit ethical, economic or practical walls, expanding HCI's ability to explore complex design spaces while maintaining research integrity. 

\subsection{Limitations}
This research acknowledges methodological limitations that reflect ongoing debates within HCI. Our speculative design approach deliberately diverges from empirical paradigms - a choice sometimes critiqued as lacking the rigor of ethnographic or clinical methods \citep{lindley2016peer}. While traditional approaches would require medical expert involvement, extensive user testing, and statistical validation, we contend that HCI scholarship benefits from methodological diversity. Rather than positioning speculative inquiry as inferior to empirical work, we demonstrate its complementary value through \textit{speculative-statistical} analysis and systematic design exploration. When properly framed and executed with intellectual rigor, speculative research offers distinct scholarly contributions that enrich HCI's knowledge base.

Technical limitations exist within specific prototypes. The t-SNE visualization exhibits inherent non-determinism and parameter sensitivity challenging reproducible analysis \cite{wattenberg2016use}. The spectrogram's sliding window approach (5-minute windows, 30-second steps) introduces boundary artifacts potentially misinterpreted as physiological transitions and may miss rapid autonomic changes during sleep transitions or acute stress responses. However, these limitations align with our speculative exploration framework - we explicitly position these as exploratory prototypes investigating visualization possibilities rather than clinically validated tools. Further, research transparency about such technical constraints while demonstrating visualization potential exemplifies how speculative design can advance understanding despite acknowledged limitations.
\section{Conclusion}
This research explored five prototypes for personal heart rate visualization using LLM-based data generation and evaluation. Our findings reveal a critical challenge: clinically accurate visualizations often become too complex and aesthetically unappealing for users. Notably, certain approaches like Poincaré plots demonstrate that heart rate data can maintain clinical validity while remaining accessible when presented through familiar visual formats, suggesting a path forward for patient-centered health visualization design. While speculative, this work provides a foundation for future empirical studies to validate these design directions and explore dynamic adaptation to user expertise, transforming heart rate data into user actionable insights.

\bibliography{sn-bibliography}


\begin{thebibliography}{84}
\ifx \bisbn   \undefined \def \bisbn  #1{ISBN #1}\fi
\ifx \binits  \undefined \def \binits#1{#1}\fi
\ifx \bauthor  \undefined \def \bauthor#1{#1}\fi
\ifx \batitle  \undefined \def \batitle#1{#1}\fi
\ifx \bjtitle  \undefined \def \bjtitle#1{#1}\fi
\ifx \bvolume  \undefined \def \bvolume#1{\textbf{#1}}\fi
\ifx \byear  \undefined \def \byear#1{#1}\fi
\ifx \bissue  \undefined \def \bissue#1{#1}\fi
\ifx \bfpage  \undefined \def \bfpage#1{#1}\fi
\ifx \blpage  \undefined \def \blpage #1{#1}\fi
\ifx \burl  \undefined \def \burl#1{\textsf{#1}}\fi
\ifx \doiurl  \undefined \def \doiurl#1{\url{https://doi.org/#1}}\fi
\ifx \betal  \undefined \def \betal{\textit{et al.}}\fi
\ifx \binstitute  \undefined \def \binstitute#1{#1}\fi
\ifx \binstitutionaled  \undefined \def \binstitutionaled#1{#1}\fi
\ifx \bctitle  \undefined \def \bctitle#1{#1}\fi
\ifx \beditor  \undefined \def \beditor#1{#1}\fi
\ifx \bpublisher  \undefined \def \bpublisher#1{#1}\fi
\ifx \bbtitle  \undefined \def \bbtitle#1{#1}\fi
\ifx \bedition  \undefined \def \bedition#1{#1}\fi
\ifx \bseriesno  \undefined \def \bseriesno#1{#1}\fi
\ifx \blocation  \undefined \def \blocation#1{#1}\fi
\ifx \bsertitle  \undefined \def \bsertitle#1{#1}\fi
\ifx \bsnm \undefined \def \bsnm#1{#1}\fi
\ifx \bsuffix \undefined \def \bsuffix#1{#1}\fi
\ifx \bparticle \undefined \def \bparticle#1{#1}\fi
\ifx \barticle \undefined \def \barticle#1{#1}\fi
\bibcommenthead
\ifx \bconfdate \undefined \def \bconfdate #1{#1}\fi
\ifx \botherref \undefined \def \botherref #1{#1}\fi
\ifx \url \undefined \def \url#1{\textsf{#1}}\fi
\ifx \bchapter \undefined \def \bchapter#1{#1}\fi
\ifx \bbook \undefined \def \bbook#1{#1}\fi
\ifx \bcomment \undefined \def \bcomment#1{#1}\fi
\ifx \oauthor \undefined \def \oauthor#1{#1}\fi
\ifx \citeauthoryear \undefined \def \citeauthoryear#1{#1}\fi
\ifx \endbibitem  \undefined \def \endbibitem {}\fi
\ifx \bconflocation  \undefined \def \bconflocation#1{#1}\fi
\ifx \arxivurl  \undefined \def \arxivurl#1{\textsf{#1}}\fi
\csname PreBibitemsHook\endcsname

\bibitem[\protect\citeauthoryear{Choe et~al.}{2014}]{choe2014understanding}
\begin{bchapter}
\bauthor{\bsnm{Choe}, \binits{E.K.}},
\bauthor{\bsnm{Lee}, \binits{N.B.}},
\bauthor{\bsnm{Lee}, \binits{B.}},
\bauthor{\bsnm{Pratt}, \binits{W.}},
\bauthor{\bsnm{Kientz}, \binits{J.A.}}:
\bctitle{Understanding quantified-selfers' practices in collecting and
  exploring personal data}.
In: \bbtitle{Proceedings of the SIGCHI Conference on Human Factors in Computing
  Systems}.
\bsertitle{CHI '14},
pp. \bfpage{1143}--\blpage{1152}.
\bpublisher{ACM}, \blocation{???}
(\byear{2014}).
\doiurl{10.1145/2556288.2557372}
\end{bchapter}
\endbibitem

\bibitem[\protect\citeauthoryear{Huang et~al.}{2015}]{huang2015personal}
\begin{barticle}
\bauthor{\bsnm{Huang}, \binits{D.}},
\bauthor{\bsnm{Tory}, \binits{M.}},
\bauthor{\bsnm{Adriel~Aseniero}, \binits{B.}},
\bauthor{\bsnm{Bartram}, \binits{L.}},
\bauthor{\bsnm{Bateman}, \binits{S.}},
\bauthor{\bsnm{Carpendale}, \binits{S.}},
\bauthor{\bsnm{Tang}, \binits{A.}},
\bauthor{\bsnm{Woodbury}, \binits{R.}}:
\batitle{Personal visualization and personal visual analytics}.
\bjtitle{IEEE Transactions on Visualization and Computer Graphics}
\bvolume{21}(\bissue{3}),
\bfpage{420}--\blpage{433}
(\byear{2015})
\doiurl{10.1109/TVCG.2014.2359887}
\end{barticle}
\endbibitem

\bibitem[\protect\citeauthoryear{Perin et~al.}{2015}]{perin2015personal}
\begin{bchapter}
\bauthor{\bsnm{Perin}, \binits{C.}},
\bauthor{\bsnm{Thudt}, \binits{A.}},
\bauthor{\bsnm{Tory}, \binits{M.}},
\bauthor{\bsnm{Willett}, \binits{W.}},
\bauthor{\bsnm{Carpendale}, \binits{S.}}:
\bctitle{Personal visualization: Exploring data in everyday life}.
In: \bbtitle{IEEE VIS 2015 Workshop},
\bconflocation{Chicago, IL, USA}
(\byear{2015}).
\bcomment{IEEE. Workshop on Personal Visualization: Exploring Data in Everyday
  Life}
\end{bchapter}
\endbibitem

\bibitem[\protect\citeauthoryear{Bent et~al.}{2020}]{bent2020investigating}
\begin{botherref}
\oauthor{\bsnm{Bent}, \binits{B.}},
\oauthor{\bsnm{Goldstein}, \binits{B.A.}},
\oauthor{\bsnm{Kibbe}, \binits{W.A.}},
\oauthor{\bsnm{Dunn}, \binits{J.P.}}:
Investigating sources of inaccuracy in wearable optical heart rate sensors.
npj Digital Medicine
\textbf{3}(18)
(2020)
\doiurl{10.1038/s41746-020-0226-6}
\end{botherref}
\endbibitem

\bibitem[\protect\citeauthoryear{Khundaqji et~al.}{2020}]{khundaqji2020smart}
\begin{barticle}
\bauthor{\bsnm{Khundaqji}, \binits{H.}},
\bauthor{\bsnm{Hing}, \binits{W.}},
\bauthor{\bsnm{Furness}, \binits{J.}},
\bauthor{\bsnm{Climstein}, \binits{M.}}, \betal:
\batitle{Smart shirts for monitoring physiological parameters: scoping review}.
\bjtitle{JMIR mHealth and uHealth}
\bvolume{8}(\bissue{5}),
\bfpage{18092}
(\byear{2020})
\end{barticle}
\endbibitem

\bibitem[\protect\citeauthoryear{Shaffer and
  Ginsberg}{2017}]{shaffer2017overview}
\begin{barticle}
\bauthor{\bsnm{Shaffer}, \binits{F.}},
\bauthor{\bsnm{Ginsberg}, \binits{J.P.}}:
\batitle{An overview of heart rate variability metrics and norms}.
\bjtitle{Frontiers in Public Health}
\bvolume{5},
\bfpage{258}
(\byear{2017})
\doiurl{10.3389/fpubh.2017.00258}
\end{barticle}
\endbibitem

\bibitem[\protect\citeauthoryear{Dong}{2024}]{Dong16092024}
\begin{barticle}
\bauthor{\bsnm{Dong}, \binits{M.}}:
\batitle{Investigating the users’ preferences of heart rate data types and
  visualizations on a smartwatch}.
\bjtitle{International Journal of Human–Computer Interaction}
\bvolume{40}(\bissue{18}),
\bfpage{5298}--\blpage{5319}
(\byear{2024})
\doiurl{10.1080/10447318.2023.2233130}
{\href{https://arxiv.org/abs/https://doi.org/10.1080/10447318.2023.2233130}{{https://doi.org/10.1080/10447318.2023.2233130}}}
\end{barticle}
\endbibitem

\bibitem[\protect\citeauthoryear{Li et~al.}{2023}]{li2023heart}
\begin{barticle}
\bauthor{\bsnm{Li}, \binits{K.}},
\bauthor{\bsnm{Cardoso}, \binits{C.}},
\bauthor{\bsnm{Moctezuma-Ramirez}, \binits{A.}},
\bauthor{\bsnm{Elgalad}, \binits{A.}},
\bauthor{\bsnm{Perin}, \binits{E.}}:
\batitle{Heart rate variability measurement through a smart wearable device:
  another breakthrough for personal health monitoring?}
\bjtitle{International journal of environmental research and public health}
\bvolume{20}(\bissue{24}),
\bfpage{7146}
(\byear{2023})
\end{barticle}
\endbibitem

\bibitem[\protect\citeauthoryear{De~Zambotti et~al.}{2024}]{de2024state}
\begin{barticle}
\bauthor{\bsnm{De~Zambotti}, \binits{M.}},
\bauthor{\bsnm{Goldstein}, \binits{C.}},
\bauthor{\bsnm{Cook}, \binits{J.}},
\bauthor{\bsnm{Menghini}, \binits{L.}},
\bauthor{\bsnm{Altini}, \binits{M.}},
\bauthor{\bsnm{Cheng}, \binits{P.}},
\bauthor{\bsnm{Robillard}, \binits{R.}}:
\batitle{State of the science and recommendations for using wearable technology
  in sleep and circadian research}.
\bjtitle{Sleep}
\bvolume{47}(\bissue{4}),
\bfpage{325}
(\byear{2024})
\end{barticle}
\endbibitem

\bibitem[\protect\citeauthoryear{Rodrigues et~al.}{2022}]{rodrigues2022hrv}
\begin{barticle}
\bauthor{\bsnm{Rodrigues}, \binits{E.}},
\bauthor{\bsnm{Lima}, \binits{D.}},
\bauthor{\bsnm{Barbosa}, \binits{P.}},
\bauthor{\bsnm{Gonzaga}, \binits{K.}},
\bauthor{\bsnm{Guerra}, \binits{R.O.}},
\bauthor{\bsnm{Pimentel}, \binits{M.}},
\bauthor{\bsnm{Barbosa}, \binits{H.}},
\bauthor{\bsnm{Maciel}, \binits{{\'A}.}}:
\batitle{Hrv monitoring using commercial wearable devices as a health indicator
  for older persons during the pandemic}.
\bjtitle{Sensors}
\bvolume{22}(\bissue{5}),
\bfpage{2001}
(\byear{2022})
\end{barticle}
\endbibitem

\bibitem[\protect\citeauthoryear{Stone et~al.}{2021}]{stone2021assessing}
\begin{barticle}
\bauthor{\bsnm{Stone}, \binits{J.D.}},
\bauthor{\bsnm{Ulman}, \binits{H.K.}},
\bauthor{\bsnm{Tran}, \binits{K.}},
\bauthor{\bsnm{Thompson}, \binits{A.G.}},
\bauthor{\bsnm{Halter}, \binits{M.D.}},
\bauthor{\bsnm{Ramadan}, \binits{J.H.}},
\bauthor{\bsnm{Stephenson}, \binits{M.}},
\bauthor{\bsnm{Finomore~Jr}, \binits{V.S.}},
\bauthor{\bsnm{Galster}, \binits{S.M.}},
\bauthor{\bsnm{Rezai}, \binits{A.R.}}, \betal:
\batitle{Assessing the accuracy of popular commercial technologies that measure
  resting heart rate and heart rate variability}.
\bjtitle{Frontiers in Sports and Active Living}
\bvolume{3},
\bfpage{585870}
(\byear{2021})
\end{barticle}
\endbibitem

\bibitem[\protect\citeauthoryear{Kim et~al.}{2018}]{kim2018stress}
\begin{barticle}
\bauthor{\bsnm{Kim}, \binits{H.-G.}},
\bauthor{\bsnm{Cheon}, \binits{E.-J.}},
\bauthor{\bsnm{Bai}, \binits{D.-S.}},
\bauthor{\bsnm{Lee}, \binits{Y.H.}},
\bauthor{\bsnm{Koo}, \binits{B.-H.}}:
\batitle{Stress and heart rate variability: a meta-analysis and review of the
  literature}.
\bjtitle{Psychiatry investigation}
\bvolume{15}(\bissue{3}),
\bfpage{235}
(\byear{2018})
\end{barticle}
\endbibitem

\bibitem[\protect\citeauthoryear{Bleecker}{2022}]{bleecker2022design}
\begin{botherref}
\oauthor{\bsnm{Bleecker}, \binits{J.}}:
Design fiction: A short essay on design, science, fact, and fiction.
Machine learning and the city: applications in architecture and urban design,
561--578
(2022)
\end{botherref}
\endbibitem

\bibitem[\protect\citeauthoryear{Tanenbaum et~al.}{2012}]{tanenbaum2012design}
\begin{bchapter}
\bauthor{\bsnm{Tanenbaum}, \binits{T.J.}},
\bauthor{\bsnm{Tanenbaum}, \binits{K.}},
\bauthor{\bsnm{Wakkary}, \binits{R.}}:
\bctitle{Design fictions}.
In: \bbtitle{Proceedings of the Sixth International Conference on Tangible,
  Embedded and Embodied Interaction},
pp. \bfpage{347}--\blpage{350}
(\byear{2012})
\end{bchapter}
\endbibitem

\bibitem[\protect\citeauthoryear{Dunne and Raby}{2013}]{dunne2013speculative}
\begin{bbook}
\bauthor{\bsnm{Dunne}, \binits{A.}},
\bauthor{\bsnm{Raby}, \binits{F.}}:
\bbtitle{Speculative Everything: Design, Fiction, and Social Dreaming}.
\bpublisher{MIT Press},
\blocation{Cambridge, MA}
(\byear{2013})
\end{bbook}
\endbibitem

\bibitem[\protect\citeauthoryear{Wong and
  Khovanskaya}{2018}]{wong2018speculative}
\begin{bchapter}
\bauthor{\bsnm{Wong}, \binits{R.Y.}},
\bauthor{\bsnm{Khovanskaya}, \binits{V.}}:
\bctitle{Speculative design in hci: From corporate imaginations to critical
  orientations}.
In: \bbtitle{New Directions in Third Wave Human-Computer Interaction: Volume 2
  - Methodologies},
pp. \bfpage{175}--\blpage{202}.
\bpublisher{Springer},
\blocation{Cham}
(\byear{2018}).
\doiurl{10.1007/978-3-319-73374-6_10}
\end{bchapter}
\endbibitem

\bibitem[\protect\citeauthoryear{Odom et~al.}{2016}]{odom2016research}
\begin{bchapter}
\bauthor{\bsnm{Odom}, \binits{W.}},
\bauthor{\bsnm{Wakkary}, \binits{R.}},
\bauthor{\bsnm{Lim}, \binits{Y.-k.}},
\bauthor{\bsnm{Desjardins}, \binits{A.}},
\bauthor{\bsnm{Hengeveld}, \binits{B.}},
\bauthor{\bsnm{Banks}, \binits{R.}}:
\bctitle{From research prototype to research product}.
In: \bbtitle{Proceedings of the 2016 CHI Conference on Human Factors in
  Computing Systems},
pp. \bfpage{2549}--\blpage{2561}
(\byear{2016})
\end{bchapter}
\endbibitem

\bibitem[\protect\citeauthoryear{Elsden et~al.}{2017}]{elsden2017speculative}
\begin{bchapter}
\bauthor{\bsnm{Elsden}, \binits{C.}},
\bauthor{\bsnm{Chatting}, \binits{D.}},
\bauthor{\bsnm{Durrant}, \binits{A.C.}},
\bauthor{\bsnm{Garbett}, \binits{A.}},
\bauthor{\bsnm{Nissen}, \binits{B.}},
\bauthor{\bsnm{Vines}, \binits{J.}},
\bauthor{\bsnm{Kirk}, \binits{D.S.}}:
\bctitle{On speculative enactments}.
In: \bbtitle{Proceedings of the 2017 CHI Conference on Human Factors in
  Computing Systems}.
\bsertitle{CHI '17},
pp. \bfpage{5386}--\blpage{5399}.
\bpublisher{ACM},
\blocation{New York, NY, USA}
(\byear{2017}).
\doiurl{10.1145/3025453.3025503}
\end{bchapter}
\endbibitem

\bibitem[\protect\citeauthoryear{Blythe et~al.}{2025}]{blythe2025artificial}
\begin{bchapter}
\bauthor{\bsnm{Blythe}, \binits{M.}},
\bauthor{\bsnm{Lindley}, \binits{S.}},
\bauthor{\bsnm{Murray-Rust}, \binits{D.}}:
\bctitle{Artificial intelligence and other speculative metaphors}.
In: \bbtitle{Proceedings of the 2025 ACM Designing Interactive Systems
  Conference},
pp. \bfpage{347}--\blpage{356}
(\byear{2025})
\end{bchapter}
\endbibitem

\bibitem[\protect\citeauthoryear{Panda}{2024}]{panda2024thought}
\begin{botherref}
\oauthor{\bsnm{Panda}, \binits{S.}}:
Thought experiments in design fiction for visualization.
arXiv preprint arXiv:2411.08621
(2024)
\end{botherref}
\endbibitem

\bibitem[\protect\citeauthoryear{Pousman et~al.}{2007}]{pousman2007casual}
\begin{bchapter}
\bauthor{\bsnm{Pousman}, \binits{Z.}},
\bauthor{\bsnm{Stasko}, \binits{J.}},
\bauthor{\bsnm{Mateas}, \binits{M.}}:
\bctitle{Casual information visualization: Depictions of data in everyday
  life},
vol. \bseriesno{13},
pp. \bfpage{1145}--\blpage{1152}.
\bpublisher{IEEE}, \blocation{???}
(\byear{2007}).
\doiurl{10.1109/TVCG.2007.70541}
\end{bchapter}
\endbibitem

\bibitem[\protect\citeauthoryear{Aseniero et~al.}{2020}]{aseniero2020activity}
\begin{bchapter}
\bauthor{\bsnm{Aseniero}, \binits{B.A.}},
\bauthor{\bsnm{Perin}, \binits{C.}},
\bauthor{\bsnm{Willett}, \binits{W.}},
\bauthor{\bsnm{Tang}, \binits{A.}},
\bauthor{\bsnm{Carpendale}, \binits{S.}}:
\bctitle{Activity river: Visualizing planned and logged personal activities for
  reflection}.
In: \bbtitle{Proceedings of the 2020 International Conference on Advanced
  Visual Interfaces},
pp. \bfpage{4}--\blpage{149}.
\bpublisher{ACM}, \blocation{???}
(\byear{2020})
\end{bchapter}
\endbibitem

\bibitem[\protect\citeauthoryear{Perin}{2021}]{perin2021students}
\begin{barticle}
\bauthor{\bsnm{Perin}, \binits{C.}}:
\batitle{What students learn with personal data physicalization}.
\bjtitle{IEEE Computer Graphics and Applications}
\bvolume{41}(\bissue{6}),
\bfpage{48}--\blpage{58}
(\byear{2021})
\doiurl{10.1109/MCG.2021.3115417}
\end{barticle}
\endbibitem

\bibitem[\protect\citeauthoryear{Choe et~al.}{2015}]{choe2015characterizing}
\begin{barticle}
\bauthor{\bsnm{Choe}, \binits{E.K.}},
\bauthor{\bsnm{Lee}, \binits{B.}},
\bauthor{\bsnm{Schraefel}, \binits{M.}}:
\batitle{Characterizing visualization insights from quantified selfers'
  personal data presentations}.
\bjtitle{IEEE Computer Graphics and Applications}
\bvolume{35}(\bissue{4}),
\bfpage{28}--\blpage{37}
(\byear{2015})
\doiurl{10.1109/MCG.2015.51}
\end{barticle}
\endbibitem

\bibitem[\protect\citeauthoryear{Rajabiyazdi
  et~al.}{2020}]{rajabiyazdi2020exploring}
\begin{bchapter}
\bauthor{\bsnm{Rajabiyazdi}, \binits{F.}},
\bauthor{\bsnm{Perin}, \binits{C.}},
\bauthor{\bsnm{Oehlberg}, \binits{L.}},
\bauthor{\bsnm{Carpendale}, \binits{S.}}:
\bctitle{Exploring the design of patient-generated data visualizations}.
In: \bbtitle{Graphics Interface Conference (GI)}
(\byear{2020}).
\bcomment{Canadian Information Processing Society}
\end{bchapter}
\endbibitem

\bibitem[\protect\citeauthoryear{Choe et~al.}{2017}]{choe2017understanding}
\begin{bchapter}
\bauthor{\bsnm{Choe}, \binits{E.K.}},
\bauthor{\bsnm{Lee}, \binits{B.}},
\bauthor{\bsnm{Zhu}, \binits{H.}},
\bauthor{\bsnm{Riche}, \binits{N.H.}},
\bauthor{\bsnm{Baur}, \binits{D.}}:
\bctitle{Understanding self-reflection: How people reflect on personal data
  through visual data exploration}.
In: \bbtitle{Proceedings of the 11th EAI International Conference on Pervasive
  Computing Technologies for Healthcare},
pp. \bfpage{173}--\blpage{182}.
\bpublisher{ACM}, \blocation{???}
(\byear{2017}).
\doiurl{10.1145/3154862.3154881}
\end{bchapter}
\endbibitem

\bibitem[\protect\citeauthoryear{Van~Wijk}{2005}]{van2005value}
\begin{bchapter}
\bauthor{\bsnm{Van~Wijk}, \binits{J.J.}}:
\bctitle{The value of visualization}.
In: \bbtitle{VIS 05. IEEE Visualization, 2005.},
pp. \bfpage{79}--\blpage{86}
(\byear{2005}).
\bcomment{IEEE}
\end{bchapter}
\endbibitem

\bibitem[\protect\citeauthoryear{Xue et~al.}{2022}]{xue2022understanding}
\begin{bchapter}
\bauthor{\bsnm{Xue}, \binits{M.}},
\bauthor{\bsnm{Liang}, \binits{R.-H.}},
\bauthor{\bsnm{Hu}, \binits{J.}},
\bauthor{\bsnm{Yu}, \binits{B.}},
\bauthor{\bsnm{Feijs}, \binits{L.}}:
\bctitle{Understanding how group workers reflect on organizational stress with
  a shared, anonymous heart rate variability data visualization}.
In: \bbtitle{CHI Conference on Human Factors in Computing Systems Extended
  Abstracts},
pp. \bfpage{1}--\blpage{7}
(\byear{2022})
\end{bchapter}
\endbibitem

\bibitem[\protect\citeauthoryear{Holzinger
  et~al.}{2013}]{holzinger2013interactive}
\begin{bchapter}
\bauthor{\bsnm{Holzinger}, \binits{A.}},
\bauthor{\bsnm{Bruschi}, \binits{M.}},
\bauthor{\bsnm{Eder}, \binits{W.}}:
\bctitle{On interactive data visualization of physiological low-cost-sensor
  data with focus on mental stress}.
In: \bbtitle{International Conference on Availability, Reliability, and
  Security},
pp. \bfpage{469}--\blpage{480}
(\byear{2013}).
\bcomment{Springer}
\end{bchapter}
\endbibitem

\bibitem[\protect\citeauthoryear{Guo et~al.}{2020}]{guo2020heat}
\begin{barticle}
\bauthor{\bsnm{Guo}, \binits{H.}},
\bauthor{\bsnm{Zhang}, \binits{W.}},
\bauthor{\bsnm{Ni}, \binits{C.}},
\bauthor{\bsnm{Cai}, \binits{Z.}},
\bauthor{\bsnm{Chen}, \binits{S.}},
\bauthor{\bsnm{Huang}, \binits{X.}}:
\batitle{Heat map visualization for electrocardiogram data analysis}.
\bjtitle{BMC Cardiovascular Disorders}
\bvolume{20}(\bissue{1}),
\bfpage{277}
(\byear{2020})
\end{barticle}
\endbibitem

\bibitem[\protect\citeauthoryear{Hsu et~al.}{2021}]{9630507}
\begin{bchapter}
\bauthor{\bsnm{Hsu}, \binits{P.-Y.}},
\bauthor{\bsnm{Hsu}, \binits{P.-H.}},
\bauthor{\bsnm{Liu}, \binits{H.-L.}},
\bauthor{\bsnm{Lin}, \binits{C.-T.}},
\bauthor{\bsnm{Chou}, \binits{H.-T.}},
\bauthor{\bsnm{Tseng}, \binits{Y.-F.}},
\bauthor{\bsnm{Lee}, \binits{T.-H.}}:
\bctitle{Star-ecg: Visualization of electrocardiograms for arrhythmia and heart
  rate variability}.
In: \bbtitle{2021 43rd Annual International Conference of the IEEE Engineering
  in Medicine \& Biology Society (EMBC)},
pp. \bfpage{2815}--\blpage{2821}
(\byear{2021}).
\doiurl{10.1109/EMBC46164.2021.9630507}
\end{bchapter}
\endbibitem

\bibitem[\protect\citeauthoryear{Meuschke et~al.}{2021}]{meuschke2021gucci}
\begin{barticle}
\bauthor{\bsnm{Meuschke}, \binits{M.}},
\bauthor{\bsnm{Niemann}, \binits{U.}},
\bauthor{\bsnm{Behrendt}, \binits{B.}},
\bauthor{\bsnm{Gutberlet}, \binits{M.}},
\bauthor{\bsnm{Preim}, \binits{B.}},
\bauthor{\bsnm{Lawonn}, \binits{K.}}:
\batitle{Gucci-guided cardiac cohort investigation of blood flow data}.
\bjtitle{IEEE Transactions on Visualization and Computer Graphics}
\bvolume{29}(\bissue{3}),
\bfpage{1876}--\blpage{1892}
(\byear{2021})
\end{barticle}
\endbibitem

\bibitem[\protect\citeauthoryear{Sadeghi et~al.}{2020}]{sadeghi2020immersive}
\begin{barticle}
\bauthor{\bsnm{Sadeghi}, \binits{A.H.}},
\bauthor{\bsnm{Bakhuis}, \binits{W.}},
\bauthor{\bsnm{Van~Schaagen}, \binits{F.}},
\bauthor{\bsnm{Oei}, \binits{F.B.}},
\bauthor{\bsnm{Bekkers}, \binits{J.A.}},
\bauthor{\bsnm{Maat}, \binits{A.P.}},
\bauthor{\bsnm{Mahtab}, \binits{E.A.}},
\bauthor{\bsnm{Bogers}, \binits{A.J.}},
\bauthor{\bsnm{Taverne}, \binits{Y.J.}}:
\batitle{Immersive 3d virtual reality imaging in planning minimally invasive
  and complex adult cardiac surgery}.
\bjtitle{European Heart Journal-Digital Health}
\bvolume{1}(\bissue{1}),
\bfpage{62}--\blpage{70}
(\byear{2020})
\end{barticle}
\endbibitem

\bibitem[\protect\citeauthoryear{Oh et~al.}{2018}]{oh2018automated}
\begin{barticle}
\bauthor{\bsnm{Oh}, \binits{S.L.}},
\bauthor{\bsnm{Ng}, \binits{E.Y.}},
\bauthor{\bsnm{San~Tan}, \binits{R.}},
\bauthor{\bsnm{Acharya}, \binits{U.R.}}:
\batitle{Automated diagnosis of arrhythmia using combination of cnn and lstm
  techniques with variable length heart beats}.
\bjtitle{Computers in biology and medicine}
\bvolume{102},
\bfpage{278}--\blpage{287}
(\byear{2018})
\end{barticle}
\endbibitem

\bibitem[\protect\citeauthoryear{Henriques
  et~al.}{2020}]{henriques2020nonlinear}
\begin{barticle}
\bauthor{\bsnm{Henriques}, \binits{T.}},
\bauthor{\bsnm{Ribeiro}, \binits{M.}},
\bauthor{\bsnm{Teixeira}, \binits{A.}},
\bauthor{\bsnm{Castro}, \binits{L.}},
\bauthor{\bsnm{Antunes}, \binits{L.}},
\bauthor{\bsnm{Costa-Santos}, \binits{C.}}:
\batitle{Nonlinear methods most applied to heart-rate time series: a review}.
\bjtitle{Entropy}
\bvolume{22}(\bissue{3}),
\bfpage{309}
(\byear{2020})
\end{barticle}
\endbibitem

\bibitem[\protect\citeauthoryear{Hassib et~al.}{2017}]{hassib2017heartchat}
\begin{bchapter}
\bauthor{\bsnm{Hassib}, \binits{M.}},
\bauthor{\bsnm{Buschek}, \binits{D.}},
\bauthor{\bsnm{Wo{\'z}niak}, \binits{P.W.}},
\bauthor{\bsnm{Alt}, \binits{F.}}:
\bctitle{Heartchat: Heart rate augmented mobile chat to support empathy and
  awareness}.
In: \bbtitle{Proceedings of the 2017 CHI Conference on Human Factors in
  Computing Systems},
pp. \bfpage{2239}--\blpage{2251}.
\bpublisher{ACM}, \blocation{???}
(\byear{2017})
\end{bchapter}
\endbibitem

\bibitem[\protect\citeauthoryear{Tadas and Coyle}{2020}]{tadas2020barriers}
\begin{barticle}
\bauthor{\bsnm{Tadas}, \binits{S.}},
\bauthor{\bsnm{Coyle}, \binits{D.}}:
\batitle{Barriers to and facilitators of technology in cardiac rehabilitation:
  Systematic qualitative grounded theory review}.
\bjtitle{JMIR mHealth and uHealth}
\bvolume{8}(\bissue{11}),
\bfpage{18686}
(\byear{2020})
\end{barticle}
\endbibitem

\bibitem[\protect\citeauthoryear{Rook et~al.}{2025}]{rook2025heart}
\begin{botherref}
\oauthor{\bsnm{Rook}, \binits{C.}},
\oauthor{\bsnm{Panda}, \binits{S.}},
\oauthor{\bsnm{Sice}, \binits{P.}},
\oauthor{\bsnm{Fenwick}, \binits{A.}},
\oauthor{\bsnm{Hodgson}, \binits{D.}},
\oauthor{\bsnm{Sice}, \binits{M.D.}},
\oauthor{\bsnm{Heckels}, \binits{D.O.}},
\oauthor{\bsnm{Elvin}, \binits{G.}}:
Heart garden: Visualizing heart rate data using the metaphor of a garden.
International Journal of Human--Computer Interaction,
1--17
(2025)
\end{botherref}
\endbibitem

\bibitem[\protect\citeauthoryear{Shahid et~al.}{2025}]{shahid2025diagnostic}
\begin{barticle}
\bauthor{\bsnm{Shahid}, \binits{S.}},
\bauthor{\bsnm{Iqbal}, \binits{M.}},
\bauthor{\bsnm{Saeed}, \binits{H.}},
\bauthor{\bsnm{Hira}, \binits{S.}},
\bauthor{\bsnm{Batool}, \binits{A.}},
\bauthor{\bsnm{Khalid}, \binits{S.}},
\bauthor{\bsnm{Tahirkheli}, \binits{N.K.}}:
\batitle{Diagnostic accuracy of apple watch electrocardiogram for atrial
  fibrillation: a systematic review and meta-analysis}.
\bjtitle{JACC: Advances}
\bvolume{4}(\bissue{2}),
\bfpage{101538}
(\byear{2025})
\end{barticle}
\endbibitem

\bibitem[\protect\citeauthoryear{Elvas et~al.}{2025}]{elvas2025role}
\begin{barticle}
\bauthor{\bsnm{Elvas}, \binits{L.B.}},
\bauthor{\bsnm{Almeida}, \binits{A.}},
\bauthor{\bsnm{Ferreira}, \binits{J.C.}}:
\batitle{The role of ai in cardiovascular event monitoring and early detection:
  Scoping literature review}.
\bjtitle{JMIR Medical Informatics}
\bvolume{13}(\bissue{1}),
\bfpage{64349}
(\byear{2025})
\end{barticle}
\endbibitem

\bibitem[\protect\citeauthoryear{Teo et~al.}{2024}]{teo2024federated}
\begin{botherref}
\oauthor{\bsnm{Teo}, \binits{Z.L.}},
\oauthor{\bsnm{Jin}, \binits{L.}},
\oauthor{\bsnm{Liu}, \binits{N.}},
\oauthor{\bsnm{Li}, \binits{S.}},
\oauthor{\bsnm{Miao}, \binits{D.}},
\oauthor{\bsnm{Zhang}, \binits{X.}},
\oauthor{\bsnm{Ng}, \binits{W.Y.}},
\oauthor{\bsnm{Tan}, \binits{T.F.}},
\oauthor{\bsnm{Lee}, \binits{D.M.}},
\oauthor{\bsnm{Chua}, \binits{K.J.}}, et al.:
Federated machine learning in healthcare: A systematic review on clinical
  applications and technical architecture.
Cell Reports Medicine
\textbf{5}(2)
(2024)
\end{botherref}
\endbibitem

\bibitem[\protect\citeauthoryear{Zimmerman
  et~al.}{2007}]{zimmerman2007research}
\begin{bchapter}
\bauthor{\bsnm{Zimmerman}, \binits{J.}},
\bauthor{\bsnm{Forlizzi}, \binits{J.}},
\bauthor{\bsnm{Evenson}, \binits{S.}}:
\bctitle{Research through design as a method for interaction design research in
  hci}.
In: \bbtitle{Proceedings of the SIGCHI Conference on Human Factors in Computing
  Systems}.
\bsertitle{CHI '07},
pp. \bfpage{493}--\blpage{502}.
\bpublisher{ACM}, \blocation{???}
(\byear{2007}).
\doiurl{10.1145/1240624.1240704}
\end{bchapter}
\endbibitem

\bibitem[\protect\citeauthoryear{Buchanan}{1992}]{buchanan1992wicked}
\begin{barticle}
\bauthor{\bsnm{Buchanan}, \binits{R.}}:
\batitle{Wicked problems in design thinking}.
\bjtitle{Design issues}
\bvolume{8}(\bissue{2}),
\bfpage{5}--\blpage{21}
(\byear{1992})
\end{barticle}
\endbibitem

\bibitem[\protect\citeauthoryear{H{\"o}{\"o}k and
  L{\"o}wgren}{2012}]{hook2012strong}
\begin{barticle}
\bauthor{\bsnm{H{\"o}{\"o}k}, \binits{K.}},
\bauthor{\bsnm{L{\"o}wgren}, \binits{J.}}:
\batitle{Strong concepts: Intermediate-level knowledge in interaction design
  research}.
\bjtitle{ACM Transactions on Computer-Human Interaction (TOCHI)}
\bvolume{19}(\bissue{3}),
\bfpage{1}--\blpage{18}
(\byear{2012})
\end{barticle}
\endbibitem

\bibitem[\protect\citeauthoryear{Gaver}{2012}]{gaver2012annotated}
\begin{botherref}
\oauthor{\bsnm{Gaver}, \binits{W.}}:
What should we expect from research through design?,
937--946
(2012)
\doiurl{10.1145/2207676.2208538}
\end{botherref}
\endbibitem

\bibitem[\protect\citeauthoryear{Koskinen et~al.}{2011}]{koskinen2011design}
\begin{bchapter}
\bauthor{\bsnm{Koskinen}, \binits{I.}},
\bauthor{\bsnm{Zimmerman}, \binits{J.}},
\bauthor{\bsnm{Binder}, \binits{T.}},
\bauthor{\bsnm{Redstrom}, \binits{J.}},
\bauthor{\bsnm{Wensveen}, \binits{S.}}:
\bctitle{Design research through practice: From the lab, field, and showroom}.
\bpublisher{Morgan Kaufmann}, \blocation{???}
(\byear{2011})
\end{bchapter}
\endbibitem

\bibitem[\protect\citeauthoryear{Dunne and Raby}{2024}]{dunne2024speculative}
\begin{bbook}
\bauthor{\bsnm{Dunne}, \binits{A.}},
\bauthor{\bsnm{Raby}, \binits{F.}}:
\bbtitle{Speculative Everything, With a New Preface by the Authors: Design,
  Fiction, and Social Dreaming}.
\bpublisher{MIT press}, \blocation{???}
(\byear{2024})
\end{bbook}
\endbibitem

\bibitem[\protect\citeauthoryear{Auger}{2013}]{auger2013speculative}
\begin{barticle}
\bauthor{\bsnm{Auger}, \binits{J.}}:
\batitle{Speculative design: crafting the speculation}.
\bjtitle{Digital Creativity}
\bvolume{24}(\bissue{1}),
\bfpage{11}--\blpage{35}
(\byear{2013})
\end{barticle}
\endbibitem

\bibitem[\protect\citeauthoryear{Pierce et~al.}{2015}]{pierce2015expanding}
\begin{bchapter}
\bauthor{\bsnm{Pierce}, \binits{J.}},
\bauthor{\bsnm{Sengers}, \binits{P.}},
\bauthor{\bsnm{Hirsch}, \binits{T.}},
\bauthor{\bsnm{Jenkins}, \binits{T.}},
\bauthor{\bsnm{Gaver}, \binits{W.}},
\bauthor{\bsnm{DiSalvo}, \binits{C.}}:
\bctitle{Expanding and refining design and criticality in hci}.
In: \bbtitle{Proceedings of the 33rd Annual ACM Conference on Human Factors in
  Computing Systems},
pp. \bfpage{2083}--\blpage{2092}
(\byear{2015})
\end{bchapter}
\endbibitem

\bibitem[\protect\citeauthoryear{Fallman}{2003}]{fallman2003design}
\begin{bchapter}
\bauthor{\bsnm{Fallman}, \binits{D.}}:
\bctitle{Design-oriented human-computer interaction}.
In: \bbtitle{Proceedings of the SIGCHI Conference on Human Factors in Computing
  Systems}.
\bsertitle{CHI '03},
pp. \bfpage{225}--\blpage{232}.
\bpublisher{ACM}, \blocation{???}
(\byear{2003}).
\doiurl{10.1145/642611.642652}
\end{bchapter}
\endbibitem

\bibitem[\protect\citeauthoryear{Greenberg and
  Buxton}{2008}]{greenberg2008usability}
\begin{bchapter}
\bauthor{\bsnm{Greenberg}, \binits{S.}},
\bauthor{\bsnm{Buxton}, \binits{B.}}:
\bctitle{Usability evaluation considered harmful (some of the time)}.
In: \bbtitle{Proceedings of the SIGCHI Conference on Human Factors in Computing
  Systems},
pp. \bfpage{111}--\blpage{120}
(\byear{2008})
\end{bchapter}
\endbibitem

\bibitem[\protect\citeauthoryear{Bardzell et~al.}{2012}]{bardzell2012critical}
\begin{bchapter}
\bauthor{\bsnm{Bardzell}, \binits{S.}},
\bauthor{\bsnm{Bardzell}, \binits{J.}},
\bauthor{\bsnm{Forlizzi}, \binits{J.}},
\bauthor{\bsnm{Zimmerman}, \binits{J.}},
\bauthor{\bsnm{Antanitis}, \binits{J.}}:
\bctitle{Critical design and critical theory: the challenge of designing for
  provocation}.
In: \bbtitle{Proceedings of the Designing Interactive Systems Conference},
pp. \bfpage{288}--\blpage{297}
(\byear{2012})
\end{bchapter}
\endbibitem

\bibitem[\protect\citeauthoryear{Gaver}{2012}]{gaver2012should}
\begin{bchapter}
\bauthor{\bsnm{Gaver}, \binits{W.}}:
\bctitle{What should we expect from research through design?}
In: \bbtitle{Proceedings of the SIGCHI Conference on Human Factors in Computing
  Systems},
pp. \bfpage{937}--\blpage{946}
(\byear{2012})
\end{bchapter}
\endbibitem

\bibitem[\protect\citeauthoryear{Hämäläinen
  et~al.}{2023}]{hamalainen2023evaluating}
\begin{bchapter}
\bauthor{\bsnm{Hämäläinen}, \binits{P.}},
\bauthor{\bsnm{Tavast}, \binits{M.}},
\bauthor{\bsnm{Kunnari}, \binits{A.}}:
\bctitle{Evaluating large language models in generating synthetic hci research
  data: a case study}.
In: \bbtitle{Proceedings of the 2023 CHI Conference on Human Factors in
  Computing Systems}.
\bsertitle{CHI '23},
pp. \bfpage{1}--\blpage{19}.
\bpublisher{ACM}, \blocation{???}
(\byear{2023}).
\doiurl{10.1145/3544548.3580688}
\end{bchapter}
\endbibitem

\bibitem[\protect\citeauthoryear{Kim et~al.}{2024}]{kim2024health}
\begin{botherref}
\oauthor{\bsnm{Kim}, \binits{Y.}},
\oauthor{\bsnm{Xu}, \binits{X.}},
\oauthor{\bsnm{McDuff}, \binits{D.}},
\oauthor{\bsnm{Breazeal}, \binits{C.}},
\oauthor{\bsnm{Park}, \binits{H.W.}}:
Health-llm: Large language models for health prediction via wearable sensor
  data.
arXiv preprint arXiv:2401.06866
(2024)
\end{botherref}
\endbibitem

\bibitem[\protect\citeauthoryear{Schuller
  et~al.}{2024}]{schuller2024generating}
\begin{bchapter}
\bauthor{\bsnm{Schuller}, \binits{A.}},
\bauthor{\bsnm{Janssen}, \binits{D.}},
\bauthor{\bsnm{Blumenr{\"o}ther}, \binits{J.}},
\bauthor{\bsnm{Probst}, \binits{T.M.}},
\bauthor{\bsnm{Schmidt}, \binits{M.}},
\bauthor{\bsnm{Kumar}, \binits{C.}}:
\bctitle{Generating personas using llms and assessing their viability}.
In: \bbtitle{Extended Abstracts of the CHI Conference on Human Factors in
  Computing Systems},
pp. \bfpage{1}--\blpage{7}
(\byear{2024})
\end{bchapter}
\endbibitem

\bibitem[\protect\citeauthoryear{Panda}{2024}]{panda2024llms}
\begin{botherref}
\oauthor{\bsnm{Panda}, \binits{S.}}:
Llms' ways of seeing user personas.
arXiv preprint arXiv:2409.14858
(2024)
\end{botherref}
\endbibitem

\bibitem[\protect\citeauthoryear{Pan et~al.}{2025}]{pan2025agentcoord}
\begin{botherref}
\oauthor{\bsnm{Pan}, \binits{B.}},
\oauthor{\bsnm{Lu}, \binits{J.}},
\oauthor{\bsnm{Wang}, \binits{K.}},
\oauthor{\bsnm{Zheng}, \binits{L.}},
\oauthor{\bsnm{Wen}, \binits{Z.}},
\oauthor{\bsnm{Feng}, \binits{Y.}},
\oauthor{\bsnm{Zhu}, \binits{M.}},
\oauthor{\bsnm{Chen}, \binits{W.}}:
Agentcoord: Visually exploring coordination strategy for llm-based multi-agent
  collaboration.
Computers \& Graphics,
104338
(2025)
\end{botherref}
\endbibitem

\bibitem[\protect\citeauthoryear{Anthropic}{2025}]{anthropic2025claude}
\begin{botherref}
\oauthor{\bsnm{Anthropic}}:
Claude {4.1} [Large language model].
Accessed: 2025-09-10
(2025).
\url{https://www.anthropic.com}
\end{botherref}
\endbibitem

\bibitem[\protect\citeauthoryear{Lan et~al.}{2025}]{lan2025more}
\begin{bchapter}
\bauthor{\bsnm{Lan}, \binits{X.}},
\bauthor{\bsnm{Wang}, \binits{Y.}},
\bauthor{\bsnm{Peng}, \binits{L.}},
\bauthor{\bsnm{Ma}, \binits{X.}}:
\bctitle{More than beautiful: Exploring design features, practical
  perspectives, and implications of artistic data visualization}.
In: \bbtitle{2025 IEEE 18th Pacific Visualization Conference (PacificVis)},
pp. \bfpage{329}--\blpage{339}
(\byear{2025}).
\bcomment{IEEE}
\end{bchapter}
\endbibitem

\bibitem[\protect\citeauthoryear{Turchioe
  et~al.}{2019}]{turchioe2019systematic}
\begin{barticle}
\bauthor{\bsnm{Turchioe}, \binits{M.R.}},
\bauthor{\bsnm{Myers}, \binits{A.}},
\bauthor{\bsnm{Isaac}, \binits{S.}},
\bauthor{\bsnm{Baik}, \binits{D.}},
\bauthor{\bsnm{Grossman}, \binits{L.V.}},
\bauthor{\bsnm{Ancker}, \binits{J.S.}},
\bauthor{\bsnm{Mamykina}, \binits{L.}}:
\batitle{A systematic review of patient-facing visualizations of personal
  health data}.
\bjtitle{Applied Clinical Informatics}
\bvolume{10}(\bissue{4}),
\bfpage{751}--\blpage{770}
(\byear{2019})
\doiurl{10.4338/ACI-2019-06-R-0081}
\end{barticle}
\endbibitem

\bibitem[\protect\citeauthoryear{Chang et~al.}{2008}]{chang2008personas}
\begin{bchapter}
\bauthor{\bsnm{Chang}, \binits{Y.-n.}},
\bauthor{\bsnm{Lim}, \binits{Y.-k.}},
\bauthor{\bsnm{Stolterman}, \binits{E.}}:
\bctitle{Personas: from theory to practices}.
In: \bbtitle{Proceedings of the 5th Nordic Conference on Human-computer
  Interaction: Building Bridges},
pp. \bfpage{439}--\blpage{442}
(\byear{2008})
\end{bchapter}
\endbibitem

\bibitem[\protect\citeauthoryear{He et~al.}{2023}]{he2023beauvis}
\begin{barticle}
\bauthor{\bsnm{He}, \binits{T.}},
\bauthor{\bsnm{Isenberg}, \binits{P.}},
\bauthor{\bsnm{Dachselt}, \binits{R.}},
\bauthor{\bsnm{Isenberg}, \binits{T.}}:
\batitle{Beauvis: A validated scale for measuring the aesthetic pleasure of
  visual representations}.
\bjtitle{IEEE Transactions on Visualization and Computer Graphics}
\bvolume{29}(\bissue{1}),
\bfpage{363}--\blpage{373}
(\byear{2023})
\end{barticle}
\endbibitem

\bibitem[\protect\citeauthoryear{Cabouat et~al.}{2024}]{cabouat2024previs}
\begin{botherref}
\oauthor{\bsnm{Cabouat}, \binits{A.-F.}},
\oauthor{\bsnm{He}, \binits{T.}},
\oauthor{\bsnm{Isenberg}, \binits{P.}},
\oauthor{\bsnm{Isenberg}, \binits{T.}}:
Previs: Perceived readability evaluation for visualizations.
IEEE Transactions on Visualization and Computer Graphics
(2024)
\end{botherref}
\endbibitem

\bibitem[\protect\citeauthoryear{Elmqvist and Yi}{2012}]{elmqvist2012patterns}
\begin{bchapter}
\bauthor{\bsnm{Elmqvist}, \binits{N.}},
\bauthor{\bsnm{Yi}, \binits{J.S.}}:
\bctitle{Patterns for visualization evaluation}.
In: \bbtitle{Proceedings of the 2012 BELIV Workshop: Beyond Time and
  Errors-Novel Evaluation Methods for Visualization},
pp. \bfpage{1}--\blpage{8}
(\byear{2012})
\end{bchapter}
\endbibitem

\bibitem[\protect\citeauthoryear{Marwan et~al.}{2007}]{marwan2007recurrence}
\begin{barticle}
\bauthor{\bsnm{Marwan}, \binits{N.}},
\bauthor{\bsnm{Romano}, \binits{M.C.}},
\bauthor{\bsnm{Thiel}, \binits{M.}},
\bauthor{\bsnm{Kurths}, \binits{J.}}:
\batitle{Recurrence plots for the analysis of complex systems}.
\bjtitle{Physics Reports}
\bvolume{438}(\bissue{5-6}),
\bfpage{237}--\blpage{329}
(\byear{2007})
\end{barticle}
\endbibitem

\bibitem[\protect\citeauthoryear{Malik et~al.}{1996}]{malik1996heart}
\begin{barticle}
\bauthor{\bsnm{Malik}, \binits{M.}},
\bauthor{\bsnm{Bigger}, \binits{J.T.}},
\bauthor{\bsnm{Camm}, \binits{A.J.}},
\bauthor{\bsnm{Kleiger}, \binits{R.E.}},
\bauthor{\bsnm{Malliani}, \binits{A.}},
\bauthor{\bsnm{Moss}, \binits{A.J.}},
\bauthor{\bsnm{Schwartz}, \binits{P.J.}}:
\batitle{Heart rate variability: standards of measurement, physiological
  interpretation and clinical use}.
\bjtitle{Circulation}
\bvolume{93}(\bissue{5}),
\bfpage{1043}--\blpage{1065}
(\byear{1996})
\end{barticle}
\endbibitem

\bibitem[\protect\citeauthoryear{Thayer et~al.}{2009}]{thayer2012heart}
\begin{barticle}
\bauthor{\bsnm{Thayer}, \binits{J.F.}},
\bauthor{\bsnm{Hansen}, \binits{A.L.}},
\bauthor{\bsnm{Saus-Rose}, \binits{E.}},
\bauthor{\bsnm{Johnsen}, \binits{B.H.}}:
\batitle{Heart rate variability, prefrontal neural function, and cognitive
  performance: the neurovisceral integration perspective on self-regulation,
  adaptation, and health}.
\bjtitle{Annals of Behavioral Medicine}
\bvolume{37}(\bissue{2}),
\bfpage{141}--\blpage{153}
(\byear{2009})
\end{barticle}
\endbibitem

\bibitem[\protect\citeauthoryear{Eckmann et~al.}{1987}]{eckmann1987recurrence}
\begin{barticle}
\bauthor{\bsnm{Eckmann}, \binits{J.-P.}},
\bauthor{\bsnm{Kamphorst}, \binits{S.O.}},
\bauthor{\bsnm{Ruelle}, \binits{D.}}:
\batitle{Recurrence plots of dynamical systems}.
\bjtitle{Europhysics Letters}
\bvolume{4}(\bissue{9}),
\bfpage{973}--\blpage{977}
(\byear{1987})
\end{barticle}
\endbibitem

\bibitem[\protect\citeauthoryear{Kemp}{2006}]{kemp2006leonardo}
\begin{bbook}
\bauthor{\bsnm{Kemp}, \binits{M.}}:
\bbtitle{Leonardo da Vinci: The Marvellous Works of Nature and Man}.
\bpublisher{Oxford University Press}, \blocation{???}
(\byear{2006})
\end{bbook}
\endbibitem

\bibitem[\protect\citeauthoryear{Cohen}{2014}]{cohen2014analyzing}
\begin{bbook}
\bauthor{\bsnm{Cohen}, \binits{M.X.}}:
\bbtitle{Analyzing Neural Time Series Data: Theory and Practice}.
\bpublisher{MIT Press}, \blocation{???}
(\byear{2014})
\end{bbook}
\endbibitem

\bibitem[\protect\citeauthoryear{Clifford et~al.}{2006}]{clifford2006advanced}
\begin{bbook}
\bauthor{\bsnm{Clifford}, \binits{G.D.}},
\bauthor{\bsnm{Azuaje}, \binits{F.}},
\bauthor{\bsnm{McSharry}, \binits{P.}}:
\bbtitle{Advanced Methods and Tools for ECG Data Analysis}.
\bpublisher{Artech House}, \blocation{???}
(\byear{2006})
\end{bbook}
\endbibitem

\bibitem[\protect\citeauthoryear{Walther and Metzger}{2005}]{van2005van}
\begin{bbook}
\bauthor{\bsnm{Walther}, \binits{I.F.}},
\bauthor{\bsnm{Metzger}, \binits{R.}}:
\bbtitle{Van Gogh: The Complete Paintings}.
\bpublisher{Taschen}, \blocation{???}
(\byear{2005})
\end{bbook}
\endbibitem

\bibitem[\protect\citeauthoryear{Brennan
  et~al.}{2001}]{brennan2001relationship}
\begin{barticle}
\bauthor{\bsnm{Brennan}, \binits{M.}},
\bauthor{\bsnm{Palaniswami}, \binits{M.}},
\bauthor{\bsnm{Kamen}, \binits{P.}}:
\batitle{Do existing measures of poincar{\'e} plot geometry reflect nonlinear
  features of heart rate variability?}
\bjtitle{IEEE Transactions on Biomedical Engineering}
\bvolume{48}(\bissue{11}),
\bfpage{1342}--\blpage{1347}
(\byear{2001})
\end{barticle}
\endbibitem

\bibitem[\protect\citeauthoryear{Van~der Maaten and
  Hinton}{2008}]{vandermaaten2008visualizing}
\begin{barticle}
\bauthor{\bsnm{Maaten}, \binits{L.}},
\bauthor{\bsnm{Hinton}, \binits{G.}}:
\batitle{Visualizing data using t-sne}.
\bjtitle{Journal of Machine Learning Research}
\bvolume{9}(\bissue{11}),
\bfpage{2579}--\blpage{2605}
(\byear{2008})
\end{barticle}
\endbibitem

\bibitem[\protect\citeauthoryear{Wattenberg et~al.}{2016}]{wattenberg2016use}
\begin{barticle}
\bauthor{\bsnm{Wattenberg}, \binits{M.}},
\bauthor{\bsnm{Vi{\'e}gas}, \binits{F.}},
\bauthor{\bsnm{Johnson}, \binits{I.}}:
\batitle{How to use t-sne effectively}.
\bjtitle{Distill}
\bvolume{1}(\bissue{10}),
\bfpage{2}
(\byear{2016})
\end{barticle}
\endbibitem

\bibitem[\protect\citeauthoryear{Tulppo et~al.}{1996}]{tulppo1996quantitative}
\begin{barticle}
\bauthor{\bsnm{Tulppo}, \binits{M.P.}},
\bauthor{\bsnm{M{\"a}kikallio}, \binits{T.H.}},
\bauthor{\bsnm{Takala}, \binits{T.E.}},
\bauthor{\bsnm{Sepp{\"a}nen}, \binits{T.}},
\bauthor{\bsnm{Huikuri}, \binits{H.V.}}:
\batitle{Quantitative beat-to-beat analysis of heart rate dynamics during
  exercise}.
\bjtitle{American Journal of Physiology-Heart and Circulatory Physiology}
\bvolume{271}(\bissue{1}),
\bfpage{244}--\blpage{252}
(\byear{1996})
\end{barticle}
\endbibitem

\bibitem[\protect\citeauthoryear{Kamen et~al.}{1996}]{kamen1996application}
\begin{barticle}
\bauthor{\bsnm{Kamen}, \binits{P.W.}},
\bauthor{\bsnm{Krum}, \binits{H.}},
\bauthor{\bsnm{Tonkin}, \binits{A.M.}}:
\batitle{Application of the poincar{\'e} plot to heart rate variability
  analysis}.
\bjtitle{Australian and New Zealand Journal of Medicine}
\bvolume{26}(\bissue{1}),
\bfpage{18}--\blpage{26}
(\byear{1996})
\end{barticle}
\endbibitem

\bibitem[\protect\citeauthoryear{Lee et~al.}{2016}]{lee2016vlat}
\begin{barticle}
\bauthor{\bsnm{Lee}, \binits{S.}},
\bauthor{\bsnm{Kim}, \binits{S.-H.}},
\bauthor{\bsnm{Kwon}, \binits{B.C.}}:
\batitle{Vlat: Development of a visualization literacy assessment test}.
\bjtitle{IEEE transactions on visualization and computer graphics}
\bvolume{23}(\bissue{1}),
\bfpage{551}--\blpage{560}
(\byear{2016})
\end{barticle}
\endbibitem

\bibitem[\protect\citeauthoryear{Pandey and Ottley}{2023}]{pandey2023mini}
\begin{bchapter}
\bauthor{\bsnm{Pandey}, \binits{S.}},
\bauthor{\bsnm{Ottley}, \binits{A.}}:
\bctitle{Mini-vlat: A short and effective measure of visualization literacy}.
In: \bbtitle{Computer Graphics Forum},
vol. \bseriesno{42},
pp. \bfpage{1}--\blpage{11}
(\byear{2023}).
\bcomment{Wiley Online Library}
\end{bchapter}
\endbibitem

\bibitem[\protect\citeauthoryear{Wobbrock and
  Kientz}{2016}]{wobbrock2016research}
\begin{barticle}
\bauthor{\bsnm{Wobbrock}, \binits{J.O.}},
\bauthor{\bsnm{Kientz}, \binits{J.A.}}:
\batitle{Research contributions in human-computer interaction}.
\bjtitle{interactions}
\bvolume{23}(\bissue{3}),
\bfpage{38}--\blpage{44}
(\byear{2016})
\end{barticle}
\endbibitem

\bibitem[\protect\citeauthoryear{Ringfort-Felner
  et~al.}{2025}]{ringfort2025quality}
\begin{bchapter}
\bauthor{\bsnm{Ringfort-Felner}, \binits{R.}},
\bauthor{\bsnm{D{\"o}rrenb{\"a}cher}, \binits{J.}},
\bauthor{\bsnm{Hassenzahl}, \binits{M.}}:
\bctitle{The quality of speculation--a scoping review}.
In: \bbtitle{Proceedings of the 2025 ACM Designing Interactive Systems
  Conference},
pp. \bfpage{2373}--\blpage{2394}
(\byear{2025})
\end{bchapter}
\endbibitem

\bibitem[\protect\citeauthoryear{Sarkar}{2025}]{sarkar2025ai}
\begin{bchapter}
\bauthor{\bsnm{Sarkar}, \binits{A.}}:
\bctitle{Ai could have written this: Birth of a classist slur in knowledge
  work}.
In: \bbtitle{Proceedings of the Extended Abstracts of the CHI Conference on
  Human Factors in Computing Systems},
pp. \bfpage{1}--\blpage{12}
(\byear{2025})
\end{bchapter}
\endbibitem

\bibitem[\protect\citeauthoryear{Lindley and Coulton}{2016}]{lindley2016peer}
\begin{bchapter}
\bauthor{\bsnm{Lindley}, \binits{J.}},
\bauthor{\bsnm{Coulton}, \binits{P.}}:
\bctitle{Peer review and design fiction: " great scott! the quotes are
  redacted"}.
In: \bbtitle{Proceedings of the 2016 CHI Conference Extended Abstracts on Human
  Factors in Computing Systems},
pp. \bfpage{583}--\blpage{595}
(\byear{2016})
\end{bchapter}
\endbibitem

\end{thebibliography}



\begin{appendices}
    

\section{Scales}

\begin{table}[h!]
\centering
\caption{BeauVis Scale Items}
\label{tab:beauvis_scale}
\begin{tabular}{@{}cp{8cm}@{}}
\toprule
\textbf{Item} & \textbf{Statement}\\
\midrule
BV1 & This visualization is Enjoyable \\
BV2 & This visualization is Likable \\
BV3 & This visualization is Pleasing\\
BV4 & This visualization is Nice\\
BV5 & This visualization is Appealing\\
\bottomrule
\end{tabular}
\end{table}

\begin{table}[h!]
\centering
\caption{PREVis Scale Items}
\label{tab:previs_scale}
\begin{tabular}{@{}cp{8cm}@{}}
\toprule
\textbf{Item} & \textbf{Statement} \\
\midrule
PV1 & It is obvious for me how to read this visualization. \\
PV2 & I can easily understand how the data is represented in this visualization. \\
PV3 & I can easily understand this visualization. \\
PV4 & I don’t find this visualization messy. \\
PV5 & I don’t find this visualization crowded. \\
PV6 & I don’t find distracting parts in this visualization. \\
PV7 & I can easily find specific elements in this visualization. \\
PV8 & I can easily identify relevant information in this visualization. \\
PV9 & I can easily retrieve information from this visualization. \\
PV10 & I find data features (for example, a minimum, or an outlier, or a trend) visible in this visualization. \\
PV11 & I can clearly see data features (for example, a minimum, or an outlier, or a trend) in this visualization. \\
\bottomrule
\end{tabular}
\end{table}

\section{The Generated Personas}

\subsection{Persona 1: Sarah Chen -- High Visualization Literacy}

\textbf{Demographics:} 34-year-old data analyst, Master's degree in Statistics, lives in San Francisco

\textbf{Mini-VLAT Level:} High (9--12 correct responses) -- Strong ability to read and interpret complex visualizations across multiple chart types

\subsubsection*{Context \& Goals}
\begin{itemize}[leftmargin=1.5cm]
    \item Competitive marathon runner who uses advanced fitness tracking devices
    \item Seeks to optimize training performance through detailed heart rate zone analysis
    \item Wants to identify patterns across training cycles and correlate with performance outcomes
\end{itemize}

\subsubsection*{Challenges}
\begin{itemize}[leftmargin=1.5cm]
    \item Often frustrated by oversimplified fitness app interfaces that do not provide enough granular data
    \item Struggles to find tools that allow custom visualization configurations
    \item Wishes she could overlay multiple data sources (sleep, nutrition, training load) with heart rate trends
\end{itemize}

\subsubsection*{Preferred Visualizations}
\begin{itemize}[leftmargin=1.5cm]
    \item Multi-layered line charts showing heart rate zones over time
    \item Scatterplots correlating heart rate variability with performance metrics
    \item Heat maps displaying training intensity across weekly/monthly cycles
    \item Complex dashboard views with multiple synchronized charts
\end{itemize}

\subsection{Persona 2: Marcus Thompson -- Moderate-High Visualization Literacy}

\textbf{Demographics:} 28-year-old personal trainer, Bachelor's degree in Exercise Science, lives in Austin

\textbf{Mini-VLAT Level:} Moderate-High (7--8 correct responses) -- Comfortable with common chart types but may struggle with more complex visualizations

\subsubsection*{Context \& Goals}
\begin{itemize}[leftmargin=1.5cm]
    \item Uses heart rate data to design training programs for clients
    \item Monitors client progress and adjusts workout intensity based on heart rate responses
    \item Wants to communicate fitness progress clearly to clients
\end{itemize}

\subsubsection*{Challenges}
\begin{itemize}[leftmargin=1.5cm]
    \item Can interpret standard charts but gets confused by novel visualization formats
    \item Sometimes misreads data when multiple variables are displayed simultaneously
    \item Needs to translate complex data into simple explanations for clients
\end{itemize}

\subsubsection*{Preferred Visualizations}
\begin{itemize}[leftmargin=1.5cm]
    \item Clean line charts showing heart rate during specific workouts
    \item Simple bar charts comparing average heart rates across different exercise types
    \item Basic trend lines showing improvement over time
    \item Color-coded zone displays (fat burn, cardio, peak zones)
\end{itemize}

\subsection{Persona 3: Linda Rodriguez -- Moderate Visualization Literacy}

\textbf{Demographics:} 52-year-old office manager, High school education, lives in Phoenix

\textbf{Mini-VLAT Level:} Moderate (5--6 correct responses) -- Can read basic charts but struggles with complex or unfamiliar visualization types

\subsubsection*{Context \& Goals}
\begin{itemize}[leftmargin=1.5cm]
    \item Recently diagnosed with hypertension, uses fitness tracker as recommended by doctor
    \item Wants to understand if daily activities and stress affect her heart rate
    \item Hopes to share meaningful data with healthcare provider during appointments
\end{itemize}

\subsubsection*{Challenges}
\begin{itemize}[leftmargin=1.5cm]
    \item Gets overwhelmed by busy interfaces with too much information
    \item Has difficulty understanding statistical concepts like averages or ranges
    \item Struggles to identify trends or patterns in data over time
    \item Often misinterprets what ``normal'' ranges mean for her age group
\end{itemize}

\subsubsection*{Preferred Visualizations}
\begin{itemize}[leftmargin=1.5cm]
    \item Simple line graphs showing daily heart rate with clear labels
    \item Basic pie charts showing time spent in different heart rate zones
    \item Traffic light color systems (green/yellow/red) for immediate understanding
    \item Large, clear numbers showing daily averages with minimal additional detail
\end{itemize}

\subsection{Persona 4: Robert Kim -- Low Visualization Literacy}

\textbf{Demographics:} 67-year-old retired teacher, High school education, lives in rural Minnesota

\textbf{Mini-VLAT Level:} Low (0--4 correct responses) -- Significant difficulty interpreting most types of data visualizations

\subsubsection*{Context \& Goals}
\begin{itemize}[leftmargin=1.5cm]
    \item Wears a basic fitness tracker given by his daughter after a mild heart episode
    \item Wants to monitor heart health as advised by cardiologist
    \item Primarily interested in knowing if his heart rate is ``okay'' or ``concerning''
\end{itemize}

\subsubsection*{Challenges}
\begin{itemize}[leftmargin=1.5cm]
    \item Cannot distinguish between different types of charts or their purposes
    \item Gets confused by any visualization with multiple data series or time periods
    \item Has difficulty understanding what the numbers and trends actually mean for his health
    \item Relies heavily on text-based summaries rather than visual displays
\end{itemize}

\subsubsection*{Preferred Visualizations}
\begin{itemize}[leftmargin=1.5cm]
    \item Large numerical displays showing current heart rate only
    \item Simple text-based alerts (``Your heart rate is normal'' vs ``Consult your doctor'')
    \item Basic progress indicators (like filling bars) rather than detailed charts
    \item Minimal visual elements with heavy reliance on written explanations
    \item Clear symbols or icons rather than abstract chart representations
\end{itemize}

\section{PREVis Subscale Results Summary}

\subsection{Visualization 1: Heatmap}
\begin{tabular}{lccccc}
\toprule
\textbf{Persona} & \textbf{Understand} & \textbf{Layout} & \textbf{DataRead} & \textbf{DataFeat} & \textbf{Profile} \\
\midrule
Sarah Chen & 7.0 & 6.0 & 7.0 & 7.0 & High VL \\
Marcus Thompson & 5.7 & 4.3 & 5.7 & 5.5 & Mod-High VL \\
Linda Rodriguez & 2.3 & 2.0 & 2.3 & 2.5 & Moderate VL \\
Robert Kim & 1.0 & 1.0 & 1.0* & N/A* & Low VL \\
\bottomrule
\end{tabular}

\subsection{Visualization 2: Recurrence Plot}
\begin{tabular}{lccccc}
\toprule
\textbf{Persona} & \textbf{Understand} & \textbf{Layout} & \textbf{DataRead} & \textbf{DataFeat} & \textbf{Profile} \\
\midrule
Sarah Chen & 5.3 & 3.7 & 5.3 & 6.0 & High VL \\
Marcus Thompson & 2.7 & 2.3 & 2.7 & 3.5 & Mod-High VL \\
Linda Rodriguez & 1.0 & 1.0 & 1.0* & N/A* & Moderate VL \\
Robert Kim & 1.0 & 1.0 & N/A* & N/A* & Low VL \\
\bottomrule
\end{tabular}

\subsection{Visualization 3: Spectrogram}
\begin{tabular}{lccccc}
\toprule
\textbf{Persona} & \textbf{Understand} & \textbf{Layout} & \textbf{DataRead} & \textbf{DataFeat} & \textbf{Profile} \\
\midrule
Sarah Chen & 5.7 & 3.7 & 5.7 & 7.0 & High VL \\
Marcus Thompson & 4.0 & 2.3 & 3.7 & 5.5 & Mod-High VL \\
Linda Rodriguez & 1.7 & 1.0 & 1.7 & 2.5 & Moderate VL \\
Robert Kim & 1.0 & 1.0 & N/A* & N/A* & Low VL \\
\bottomrule
\end{tabular}

\subsection{Visualization 4: Poincaire Plot}
\begin{tabular}{lccccc}
\toprule
\textbf{Persona} & \textbf{Understand} & \textbf{Layout} & \textbf{DataRead} & \textbf{DataFeat} & \textbf{Profile} \\
\midrule
Sarah Chen & 6.7 & 5.3 & 6.3 & 7.0 & High VL \\
Marcus Thompson & 5.0 & 3.7 & 5.0 & 5.5 & Mod-High VL \\
Linda Rodriguez & 2.3 & 2.3 & 2.3 & 3.5 & Moderate VL \\
Robert Kim & 1.0 & 1.0 & N/A* & N/A* & Low VL \\
\bottomrule
\end{tabular}

\subsection{Visualization 5: t-SNE}
\begin{tabular}{lccccc}
\toprule
\textbf{Persona} & \textbf{Understand} & \textbf{Layout} & \textbf{DataRead} & \textbf{DataFeat} & \textbf{Profile} \\
\midrule
Sarah Chen & 5.7 & 2.7 & 4.7 & 6.5 & High VL \\
Marcus Thompson & 2.7 & 1.7 & 2.7 & 3.5 & Mod-High VL \\
Linda Rodriguez & 1.0 & 1.0 & 1.0* & N/A* & Moderate VL \\
Robert Kim & 1.0 & 1.0 & N/A* & N/A* & Low VL \\
\bottomrule
\end{tabular}

\textit{*N/A indicates "I don't know/Not applicable" responses were given for these subscales}

\section{BeauVis Scale Results}
\tiny{
\subsection{Visualization 1:Heatmap}

\begin{tabular}{lcccccc}
\toprule
\textbf{Persona} & \textbf{Enjoyable} & \textbf{Likable} & \textbf{Pleasing} & \textbf{Nice} & \textbf{Appealing} & \textbf{BeauVis Score} \\
\midrule
Sarah Chen & Agree & Agree & Slightly agree & Agree & Agree & \textbf{5.8} \\
Marcus Thompson & Slightly agree & Slightly agree & Neutral & Slightly agree & Neutral & \textbf{4.4} \\
Linda Rodriguez & Slightly disagree & Disagree & Disagree & Slightly disagree & Disagree & \textbf{2.4} \\
Robert Kim & Strongly disagree & Strongly disagree & Strongly disagree & Disagree & Strongly disagree & \textbf{1.4} \\
\bottomrule
\end{tabular}

\subsection{Visualization 2: Recurrence Plot}

\begin{tabular}{lcccccc}
\toprule
\textbf{Persona} & \textbf{Enjoyable} & \textbf{Likable} & \textbf{Pleasing} & \textbf{Nice} & \textbf{Appealing} & \textbf{BeauVis Score} \\
\midrule
Sarah Chen & Slightly agree & Neutral & Neutral & Slightly agree & Neutral & \textbf{4.2} \\
Marcus Thompson & Slightly disagree & Disagree & Slightly disagree & Slightly disagree & Disagree & \textbf{2.6} \\
Linda Rodriguez & Strongly disagree & Strongly disagree & Strongly disagree & Disagree & Strongly disagree & \textbf{1.2} \\
Robert Kim & Strongly disagree & Strongly disagree & Strongly disagree & Strongly disagree & Strongly disagree & \textbf{1.0} \\
\bottomrule
\end{tabular}

\subsection{Visualization 3: Spectrogram}

\begin{tabular}{lcccccc}
\toprule
\textbf{Persona} & \textbf{Enjoyable} & \textbf{Likable} & \textbf{Pleasing} & \textbf{Nice} & \textbf{Appealing} & \textbf{BeauVis Score} \\
\midrule
Sarah Chen & Agree & Slightly agree & Slightly agree & Agree & Slightly agree & \textbf{5.4} \\
Marcus Thompson & Slightly disagree & Neutral & Slightly disagree & Neutral & Slightly disagree & \textbf{3.4} \\
Linda Rodriguez & Disagree & Disagree & Strongly disagree & Disagree & Strongly disagree & \textbf{1.8} \\
Robert Kim & Strongly disagree & Strongly disagree & Strongly disagree & Strongly disagree & Strongly disagree & \textbf{1.0} \\
\bottomrule
\end{tabular}

\subsection{Visualization 4: Poincaire Plot}

\begin{tabular}{lcccccc}
\toprule
\textbf{Persona} & \textbf{Enjoyable} & \textbf{Likable} & \textbf{Pleasing} & \textbf{Nice} & \textbf{Appealing} & \textbf{BeauVis Score} \\
\midrule
Sarah Chen & Agree & Agree & Agree & Agree & Slightly agree & \textbf{5.8} \\
Marcus Thompson & Slightly agree & Slightly agree & Agree & Slightly agree & Neutral & \textbf{4.6} \\
Linda Rodriguez & Slightly disagree & Neutral & Slightly disagree & Neutral & Slightly disagree & \textbf{3.4} \\
Robert Kim & Disagree & Disagree & Slightly disagree & Disagree & Disagree & \textbf{2.2} \\
\bottomrule
\end{tabular}

\subsection{Visualization 5: t-SNE}

\begin{tabular}{lcccccc}
\toprule
\textbf{Persona} & \textbf{Enjoyable} & \textbf{Likable} & \textbf{Pleasing} & \textbf{Nice} & \textbf{Appealing} & \textbf{BeauVis Score} \\
\midrule
Sarah Chen & Slightly agree & Neutral & Slightly disagree & Neutral & Slightly disagree & \textbf{3.6} \\
Marcus Thompson & Disagree & Slightly disagree & Disagree & Slightly disagree & Disagree & \textbf{2.4} \\
Linda Rodriguez & Strongly disagree & Strongly disagree & Strongly disagree & Disagree & Strongly disagree & \textbf{1.2} \\
Robert Kim & Strongly disagree & Strongly disagree & Strongly disagree & Strongly disagree & Strongly disagree & \textbf{1.0} \\
\bottomrule
\end{tabular}}







\end{appendices}



\end{document}